\documentclass[prb,
twocolumn,
superscriptaddress,showpacs,amsmath,amssymb]{revtex4}
\usepackage{amsfonts}
\usepackage{bm}
\usepackage{verbatim}

\usepackage{graphicx}

\begin{document}

\title{Exotic Lifshitz transitions in topological materials}

\author{G.E. Volovik}

\affiliation{Low Temperature Laboratory, Aalto University, P.O. Box 15100, FI-00076 AALTO, Finland}

\affiliation{ L.~D.~Landau Institute for
Theoretical Physics, 117940 Moscow, Russia}

\affiliation{P.N. Lebedev Physical Institute, RAS, Moscow 119991, Russia}

\date{\today}

\begin{abstract}
{ 
Topological Lifshitz transitions involve many types of topological structures in momentum and frequency-momentum spaces: Fermi surfaces, Dirac lines, Dirac and Weyl points, etc. Each of these structures has their own topological invariant ($N_1$, $N_2$,  $N_3$,  $\tilde N_3$, etc.), which supports the stability of a given topological structure. The topology of the shape of Fermi surfaces and Dirac lines, as well as the interconnection of the objects of different dimensions, lead to numerous classes of Lifshitz transitions.  The consequences of Lifshitz transitions are important in different areas of physics. The singularities emerging at the transition may enhance the transition temperature to superconductivity; the Lifshitz transition can be in the origin of the small masses of elementary particles in our Universe; the black hole horizon serves as the surface of Lifshitz transition between the vacua with type-I and  type-II Weyl points; etc.
}
\end{abstract}

\maketitle

\section{Introduction. Fermi surface, Dirac line, Weyl point}

\begin{figure}
\includegraphics[width=1.0\linewidth]{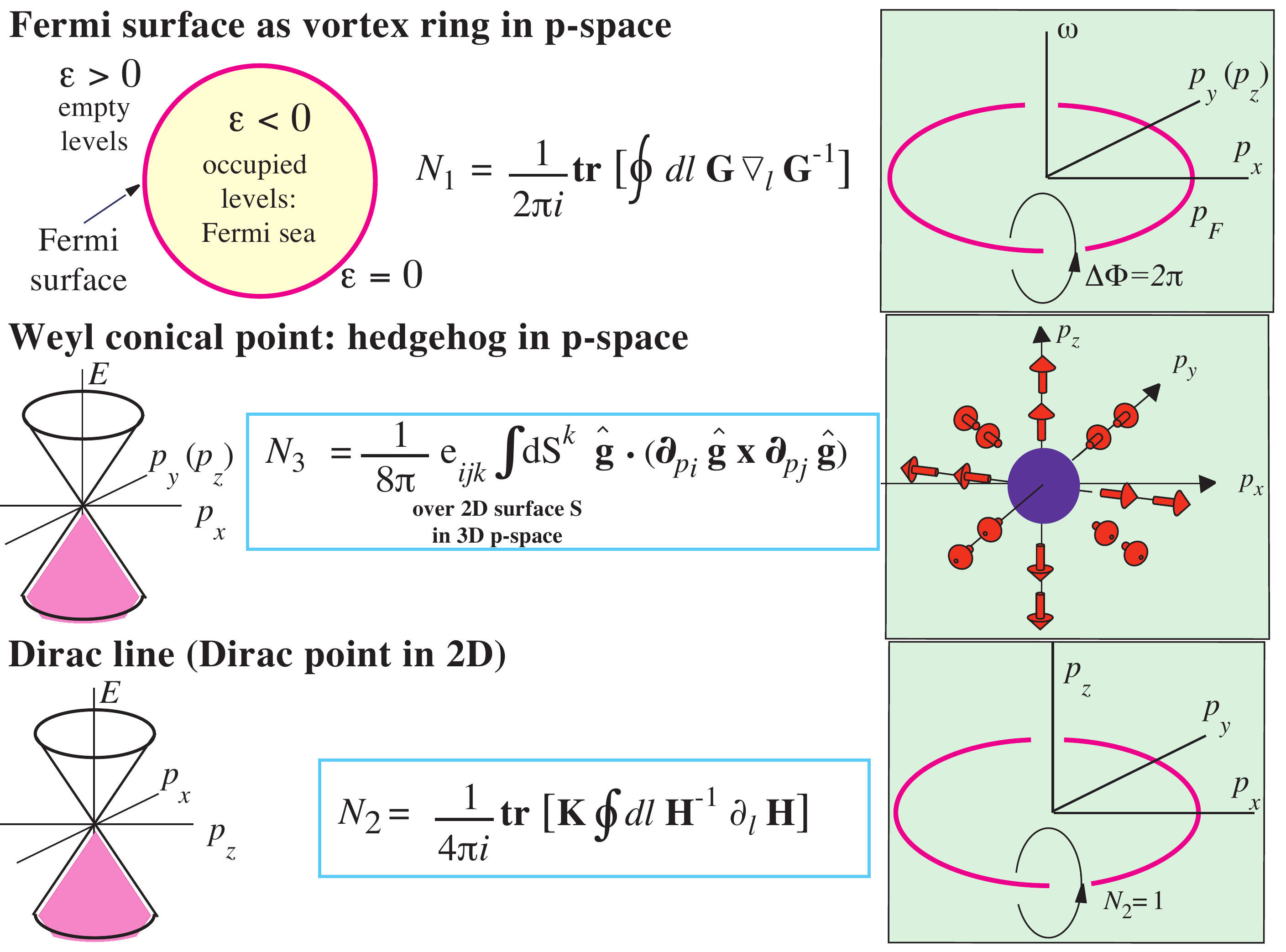}
\caption{
Topologically stable nodes in the energy spectrum of electrons in metals or fermions in general.
\\
({\it Top}): Fermi surface represents the singularity in the Green's function, which forms the vortex in the $3+1$ $({\bf p},\omega)$-space, see Sec. \ref{FSRing} and Fig. \ref{FSTopology} (in the $2+1$ $(p_x,p_y,\omega)$-space this is the vortex line). The stability of the vortex is supported by the winding number -- the integer-valued invariant $N_1$, expressed in terms of the Green's function. Lifshitz transitions, which involve the Fermi surface, are discussed in Secs. \ref{FS} and \ref{DifferentCharges}.
\\
 ({\it Middle}): Conical point in the fermionic spectrum of the Weyl materials (Weyl semimetals, chiral superfluid 
$^3$He-A, and the vacuum of Standard Model in its gapless phase, see Sec. \ref{Weyl} and Fig. \ref{WeylTopology}). The directions of spin (or of the emergent spin, isospin, pseudo-spin, etc.)  form the topological object  in momentum space -- the  hedgehog or the Berry phase monopole\cite{Volovik1987} --  described by the 
 integer-valued topological invariant $N_3$.  Lifshitz transitions, which involve the Weyl nodes, are discussed in
Secs. \ref{Weyl}, \ref{DifferentCharges} and \ref{GappedViaGapless}.
\\
 ({\it Bottom}): Dirac lines -- lines of zeroes in the energy spectrum, described by the topological invariant $N_2$. The circular line is the Dirac line in the quasiparicle spectrum in the polar phase of superfluid $^3$He,
which has been recently created in aerogel.\cite{Dmitriev2015} The same invariant $N_2$ stabilizes the point nodes in 2D materials, such as graphene, see Sec.\ref{N2}.
 }
 \label{objects}
\end{figure}

The key word in consideration of Lifshitz transitions is topology.
Following original  Lifshitz paper,\cite{ILifshitz1960}  Lifshitz transition has been viewed as a change of the topology of the Fermi surface without symmetry breaking. Later it became clear that topology of the shape  is not the only topological characterization of the Fermi surface. Fermi surface itself represents the singularity in the Green's function, which is topologically protected: it is the vortex line in the four-dimensional frequency-momentum space  in Fig. \ref{objects} ({\it top right}). The stability of the Fermi surface under interaction between the fermions is in the origin of the Fermi liquid theory developed by Landau. Moreover, the Fermi surface appeared to be one in the series of the topologically stable singularities,\cite{Horava2005,Volovik2003} 
which include in particular the Weyl point -- the hedgehog in momentum space in Fig. \ref{objects} ({\it middle}) and the Dirac line -- the vortex  line in the three-dimensional momentum space in Fig. \ref{objects} ({\it bottom}). The stability of these objects is supported by the corresponding topological invariants in momentum space or in extended frequency-momentum space.

The combination of topology of the shape of the Fermi surfaces, Fermi lines and Fermi points together with the topology, which supports the stability of these objects, and also the topology of the interconnections of the objects of different dimensions provide a large number of different types of Lifshitz transitions. Examples of Lifshitz transitions coming from the interplay of different topological objects in momentum space are discussed in Refs. \cite{Volovik2007,Volovik2017,KuangVolovik2017} and in Secs. \ref{typeII} and \ref{DifferentCharges}. 
This makes the Lifshitz transitions ubiquitous with applications to high energy physics, cosmology, black hole physics and to search for the room-$T$ superconductivity.

In particular, the Lifshitz transition may give the solution of the hierarchy problem in particle physics: 
why the masses of elementary particles in our Universe are so extremely small compared with the characteristic Planck energy scale. Indeed, when we compare the mass $\sim 10^2$ GeV  of the most heavy particle -- the top quark --  with the Planck energy $\sim 10^{19}$ GeV, we can see that the vacuum of our Universe is practically gapless. There are several topological scenarios, which may lead to the (almost) gapless vacuum.  

In one scenario the quantum vacuum belongs to the class of the gapless (massless) Weyl materials in Fig. \ref{objects} ({\it middle}), where the nodes in the spectrum of elementary particles -- the Weyl points -- are topologically
 protected,\cite{FrogNielBook,Horava2005,Volovik2003} see Sec.\ref{WeylTopologySec} and Fig. \ref{WeylTopology}. According to this scenario  the physical laws are not fundamental, but emerge in the low energy corner of the quantum vacuum, i.e. in the vicinity of the Weyl points, where the spectrum becomes linear and all the symmetries of Standard Model including Lorentz invariance and general covariance emerge from nothing. In this scenaro the  Lifshitz transition between type-I and type-II Weyl vacua take place at the black hole horizon, see Sec. \ref{BHhorizon}.

At even lower energy some of these symmetries experience spontaneous breaking, analogous to the superconducting transition. In the latter case the hierarchy problem is understood: in most superconductors the transition temperature $T_c$ is exponentially small, compared to the characteristic Fermi energy scale (analog of the Planck scale), which forces us to search for the exceptional materials with enhanced $T_c$. The role of the Lifshitz transition in the enhancement of the temperature of superconducting transition is in  Sec. \ref{FlatBand} and Sec. \ref{typeII}.

In the other scenario, the massless (gapless) vacua emerge at the Lifshits transition between the fully gapped vacua with different topological charges, see Sec.\ref{GappedViaGapless} and Fig. \ref{Fully}. The almost perfect masslessness of elementary particles in our Universe suggests that the Universe is very close to the line of the topological Lifshitz transition between the fully gapped vacua,
at which fermions necessarily become gapless, \cite{Volovik2010} see Sec. \ref{GappedViaGapless}.
This is the topological analog of the so-called Multiple Point Principle, according to which  the Universe lives at the coexistence point (line, surface, etc.) of the  first order phase transition,  where different vacua have the same energy.\cite{Nielsen2016a,Nielsen2016b,Nielsen2016c,Nielsen1997,Volovik2004}

\section{Fermi surface and Lifshitz transitions}
\label{FS}

\subsection{Fermi surface as topological object}
\label{FSRing}

\begin{figure}
\includegraphics[width=1.0\linewidth]{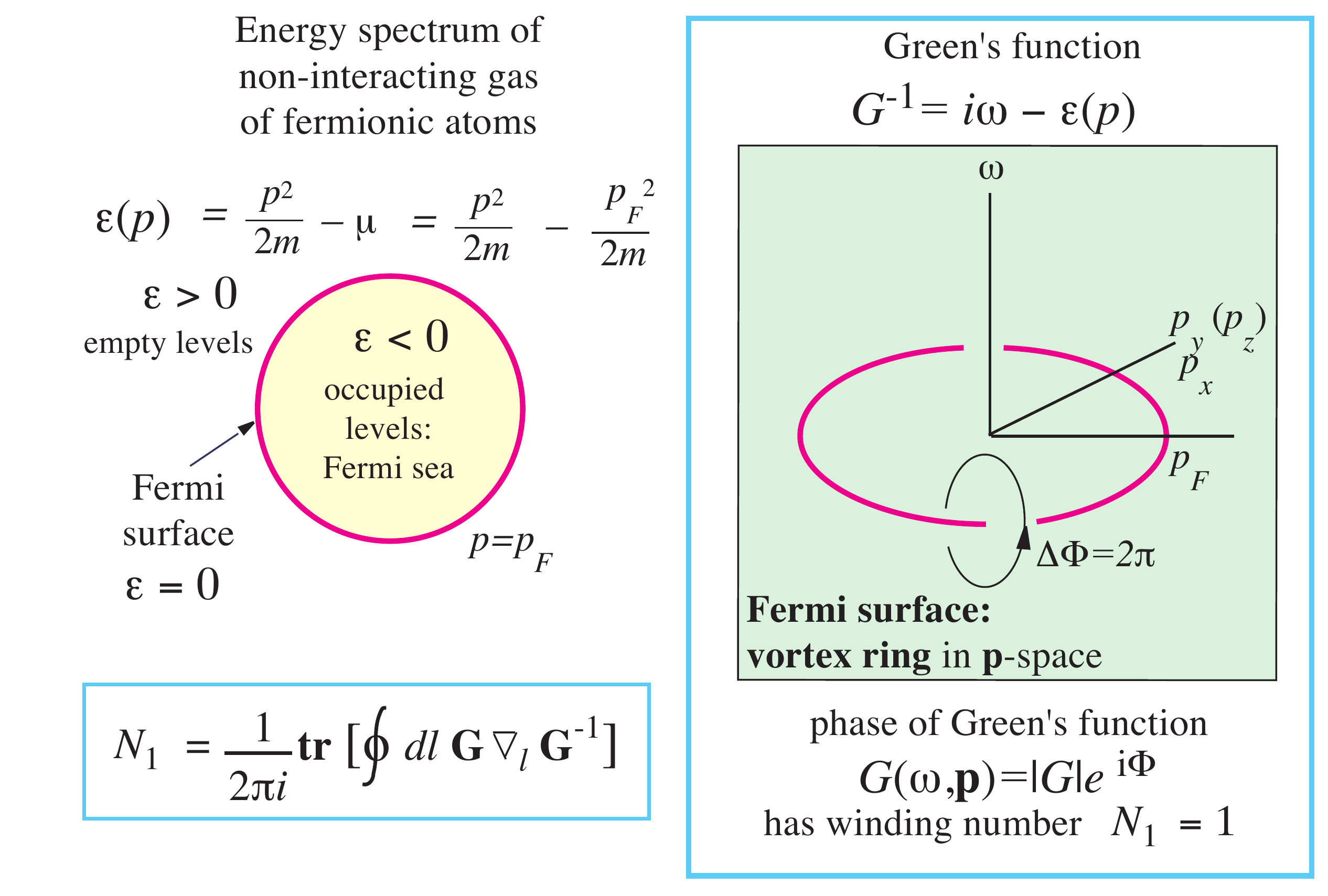}
\caption{ Fermi surface is robust to interactions because it represents the topologically stable singularity in the Green's function -- the vortex the $3+1$  $({\bf p},\omega)$-space. The stability of the vortex is supported by the winding number of the phase $\Phi$ of the Green's function $G=|G|e^ {i\Phi}$. In general the winding number is given by the integer-valued invariant $N_1$, expressed in terms of the Green's function.
 }
 \label{FSTopology}
\end{figure}

The primary topology, which is at the origin of the Lifshitz transitions,  is the topology which is reponsible for the stability of the Fermi surface.
If the Fermi surface is not stable under the electron-electron interaction, the consideration of the topology of the shape of the Fermi surface and of the corresponding Lifshitz transitions does not make much sense. To view the topological stability of the Fermi surface with respect to interactions let us start with the Green's function of an  ideal Fermi gas
in Fig. \ref{FSTopology} ({\it left}). The Fermi surface $\epsilon ({\bf p})=0$ of the noninteracting Fermi gas is the
boundary in  momentum space, which separates the occupied states with $\epsilon({\bf p}) <0$ from 
the empty states  with $\epsilon ({\bf p})>0$. The Green's function $G(\omega,{\bf p})$ with $\omega$ on imaginary axis
\begin{equation}
G^{-1}(\omega,{\bf p}) =  i\omega -\epsilon({\bf p}) \,,
\label{Propagator2}
\end{equation}
has singularity at $\omega=0$ and $\epsilon ({\bf p})=0$. In Fig. \ref{FSTopology} ({\it right}) the $p_z$ coordinate is suppressed, and the Green's function singularity forms the closed line  in the $2+1$ momentum-frequency space $(p_x,p_y,\omega)$. This line represents the vortex line, at which the phase of the Green's function $\Phi(p_x,p_y,\omega)$ has the $2\pi$ winding. As in the case of the real-space vortex in superfluids, the integer winding number provides the stability of the Fermi surface with respect to perturbations, including the interaction (if the $p_z$ component is restored, the singularity forms the vortex sheet in the $3+1$ momentum-frequency $({\bf p},\omega)$ space).

In general case, when the Green's function  has the spin, band and other indices,  the winding number can be written as the following topological invariant in terms of the Green's function:
\begin{equation}
N_1={\bf tr}~\oint_C {dl\over 2\pi i}  G(\omega,{\bf p})\partial_l
G^{-1}(\omega,{\bf p})~.
\label{InvariantForFS}
\end{equation}
Here the integral is taken over an arbitrary contour $C$ around  the
momentum-frequency  vortex sheet, and
${\bf tr}$ is the trace over all the indices.

Due to topological stability one cannot make a hole in the Fermi surface. As in the case of vortex lines, which cannot terminate in bulk,  the Fermi surface has no edges.

\subsection{Fermi surface Lifshitz transitions}

Because of the topological stability the Fermi surface may be formed even in the superconducting state. The conditions for that are multi-band structure and broken time reversal symmetry $T$ and parity $P$. \cite{Volovik1989,LiuWilczek2003,BarzykinGorkov2007,Agterberg2017,Timm2017}
These the so-called Bogolubov Fermi surfaces appear also in gapless superfluids, when the Weyl points in $^3$He-A  and Dirac nodal line in the polar phase of $^3$He are inflated to the Fermi pockets in the presence of 
superflow, which violates both $T$ and $P$ symmetries.\cite{Volovik1984,Makinen2017}

\begin{figure}
\includegraphics[width=1.0\linewidth]{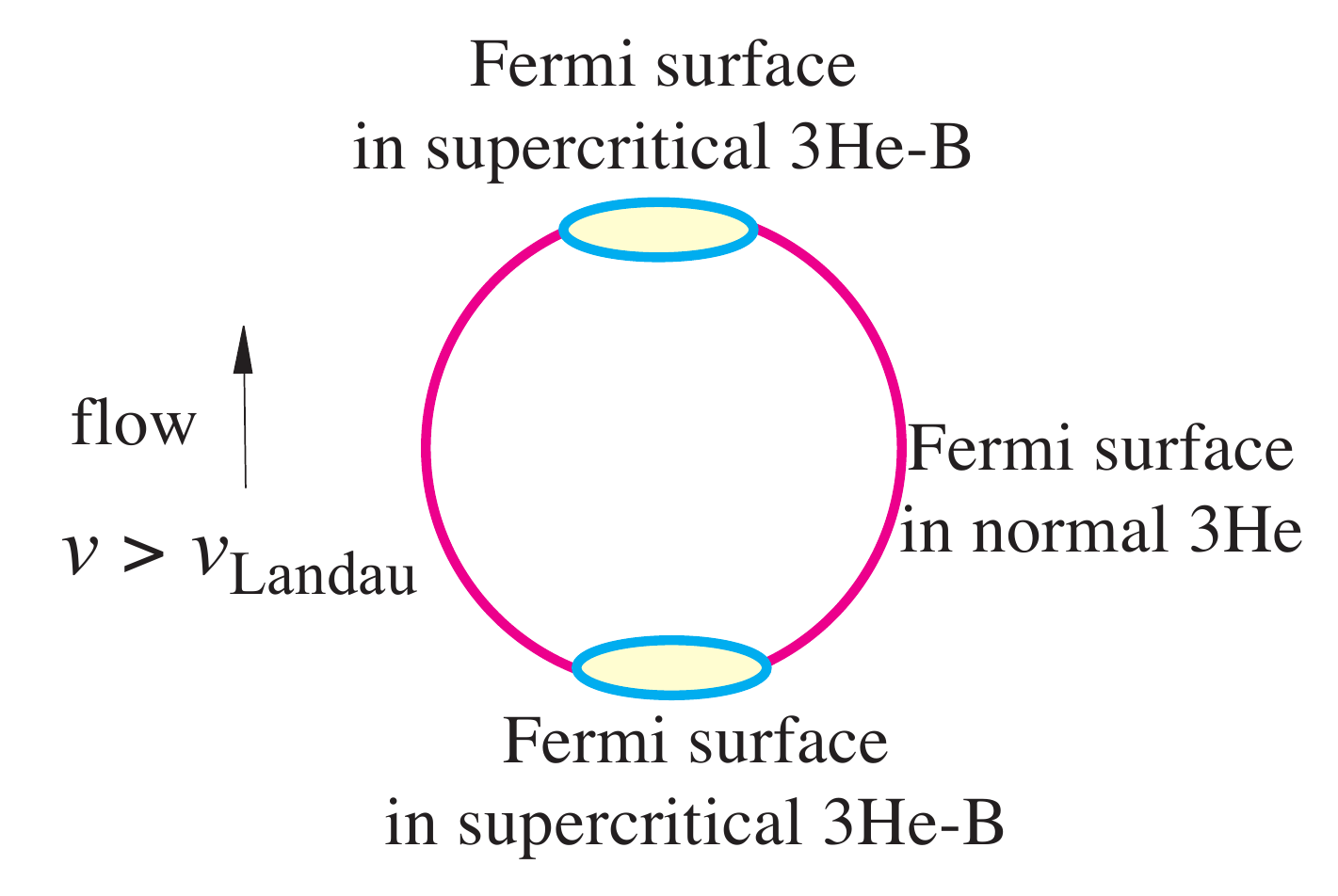}
\caption{Lifshitz transition at which the closed Fermi surfaces appear in the fully gapped superfluid $^3$He-B, when the flow velocity of the liquid with respect to the walls of container exceeds Landau critical velocity.
\cite{VollhardtMakiSchopohl1980,Volovik2003}
 }
 \label{SuperL}
\end{figure}

The Fermi surface can be also formed in the fully gapped superfluids, if the velocity of superflow exceeds Landau critical velocity.\cite{VollhardtMakiSchopohl1980,Volovik2003} The crossing of Landau velocity with formation of closed Bogoliubov Fermi surfaces is an example of one of the two transtions suggested by Lifshitz, see Fig. \ref{SuperL}.

\begin{figure}
\includegraphics[width=1.0\linewidth]{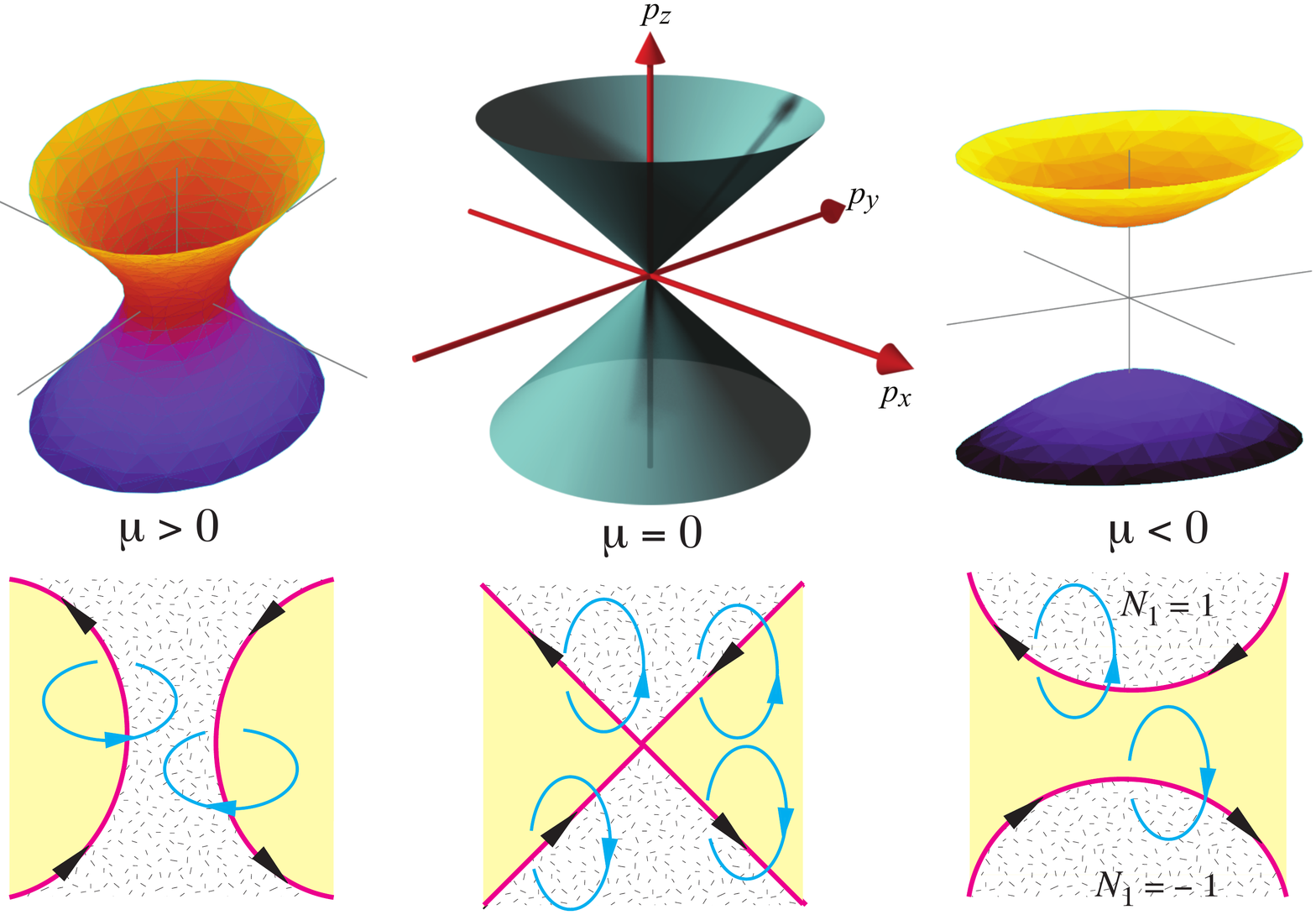}
\caption{Since the Fermi surface represents the vortex in the 3+1 $({\bf p},\omega)$ space, the Lifshits transition 
 with disruption of the neck of the Fermi surface\cite{ILifshitz1960} ({\it top}) is equivalent to the interconnection of vortices in quantum turbulence\cite{Golov2016} ({\it bottom}).
  }
 \label{Reconnection}
\end{figure}

Another original Lifshitz transition takes place when the Fermi surface crosses the stationary point of the  electronic spectrum. Near the transition the expansion of the generic spectrum has the form:\cite{ILifshitz1960}
\begin{equation}
\epsilon_{\bf p}  = ap_x^2 + bp_y^2 +cp_z^2-\mu   
\,.
\label{DisruptionTransition}
\end{equation}
For $a>0$, $b>0$, $c<0$  the transition with disruption of the neck of the Fermi surface at $\mu=0$ is in Fig. \ref{Reconnection} ({\it top}).
In terms of the vortex singularities of the Green's function in 3+1 $({\bf p},\omega)$ space, this Lifshitz transition represents the interconnection of the vortex lines in Fig. \ref{Reconnection} ({\it bottom}). In superfluids, the interconnection of the real-space vortices is an important process in the vortex turbulence.\cite{Golov2016}

\subsection{From pole of Green's function to zero}
\label{PoleZero}

\begin{figure}
\includegraphics[width=1.0\linewidth]{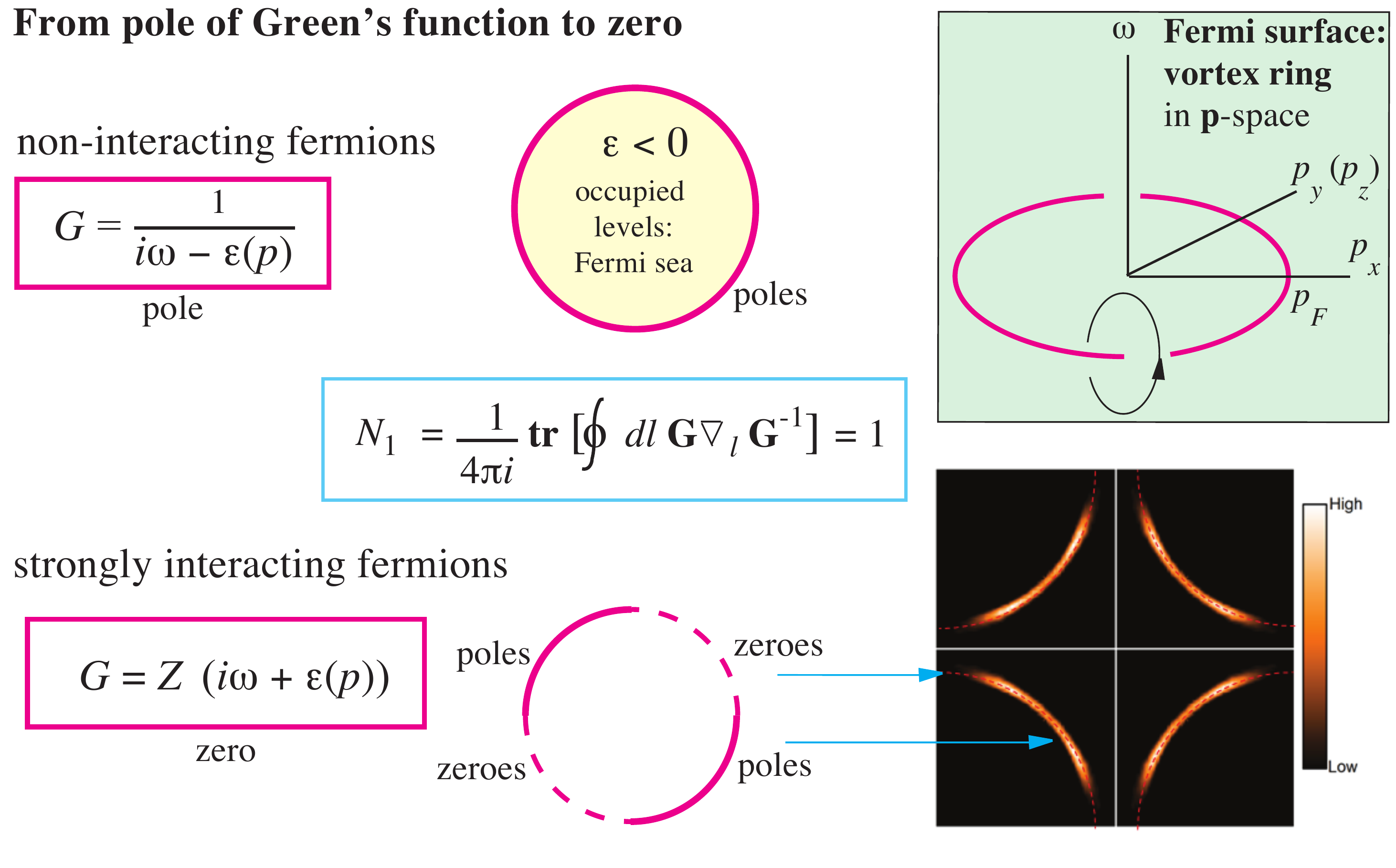}
\caption{Lifshitz transition at which the whole Fermi surface of the poles in the Green's function or part of the Fermi surface transforms to the surface of zeroes in the Green's function. The topological charge of the surface does not change at this quantum phase transition. As a result the Luttinger theorem remains valid,  \cite{Dzyaloshinskii2003,Farid2009} i.e. the particle density of interacting
fermions is equal to the volume in the momentum
space enclosed by the singular surface with the topological charge $N_1=1$.
 }
 \label{FermiArc}
\end{figure}

While in conventional Landau Fermi liquid the Green's function has a pole, for Luttinger liquid the resudue of the pole in the Green's function has singularity: the parameter $\gamma$ 
in Eq.(\ref{Zzero}) is nonzero:\cite{Voit1995}
\begin{equation}    
G=  \frac{Z} {i\omega-\epsilon({\bf p})}~~,~~ Z\propto  \left(\omega^2 +\epsilon^2({\bf p})\right)^\gamma \,.
\label{Zpole}
\end{equation}
Nevertheless the topological invariant remains the same for all $\gamma$: i.e. the
Green's function has the same topological property as the 
Green's function of conventional metal with Fermi surface at $\epsilon({\bf p})=0$.
This is the reason why the Luttinger theorem is still valid.\cite{Dzyaloshinskii2003,Farid2009}
The particle density of interacting
fermions is equal to the volume in the momentum
space enclosed by the singular surface with the topological charge $N_1=1$, irrespective of the realization of the singularity.

The suppression of residue $Z$ can be so strong, that the pole in the
Green's function  is transformed to the zero of the Green's function, which corresponds 
to the special case of $\gamma=1$, see Fig. \ref{FermiArc}:
\begin{equation}    
G\propto   i\omega+\epsilon({\bf p})\,.
\label{Zzero}
\end{equation}
This situation in particular takes place for Mott insulators,\cite{Dzyaloshinskii2003} which means that the. topology of Fermi surface is preserved even in the insulating phase, and thus  the Luttinger
theorem is still valid.\cite{Dzyaloshinskii2003,Farid2009}
Thus we can say that the transition between metals and insulators can be also viewed as a type of the zero-temperature Lifshitz transition, at which the property of the energy spectrum drastically changes without symmetry breaking. However, this quantum phase transition is not topological since the topological invariant does not change across the transition.

It is not excluded that in the so-called pseudo-gap phase of cuprate superconductors
and some other materials, see e.g.\cite{PseudGap}, the part of the Fermi surface transforms to the surface of zeroes and the Fermi arcs are formed, see Fig. \ref{FermiArc} ({\it bottom right}).

\subsection{From Fermi surface to flat band}
\label{FlatBand}

\begin{figure}
\includegraphics[width=1.0\linewidth]{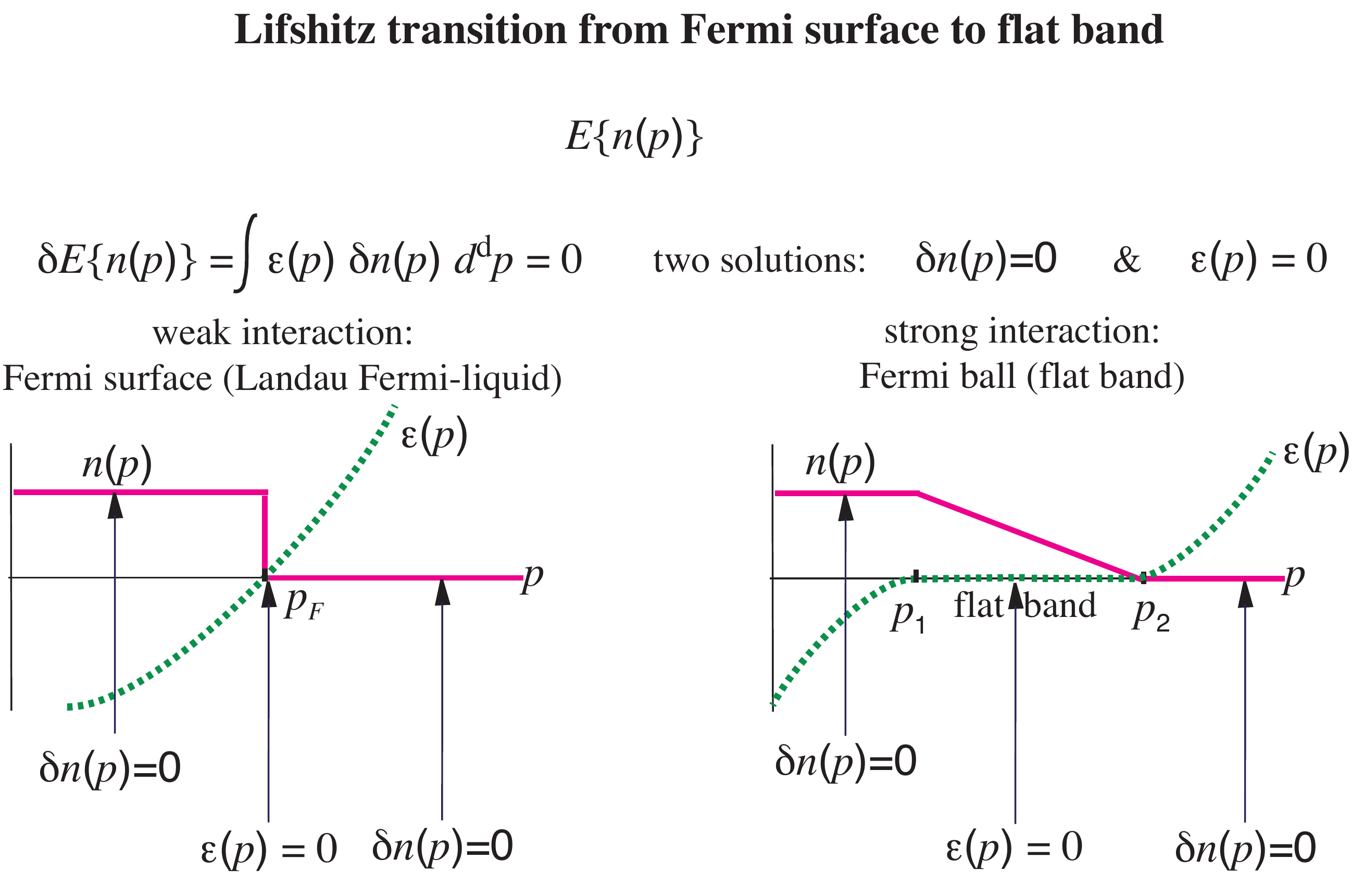}
\caption{Formation of flat band in the system of stronlgly interacting fermions in the Landau theory approach.
There are to types of the extrema of the energy functional $E\{n({\bf p})\}$: (i)   $\delta n({\bf p})=0$, which corresponds to the occupied  $n({\bf p})=1$ and free  $n({\bf p})=0$ levels.
(ii) $\epsilon({\bf p})=0$; this solution takes place when $0<n({\bf p})<1$.
On the weak interaction side, the solution (i) takes place inside and the outside the Fermi surface, while the solution (ii) corresponds to the Fermi surface. On the strong interaction side the solution$\epsilon({\bf p})=0$ is extended to the 3D band -- the flat band. Formation of the flat band from the Fermi surface occurs via the new type of Lifshitz transition.
 }
 \label{KC}
\end{figure}

\begin{figure}
\includegraphics[width=1.0\linewidth]{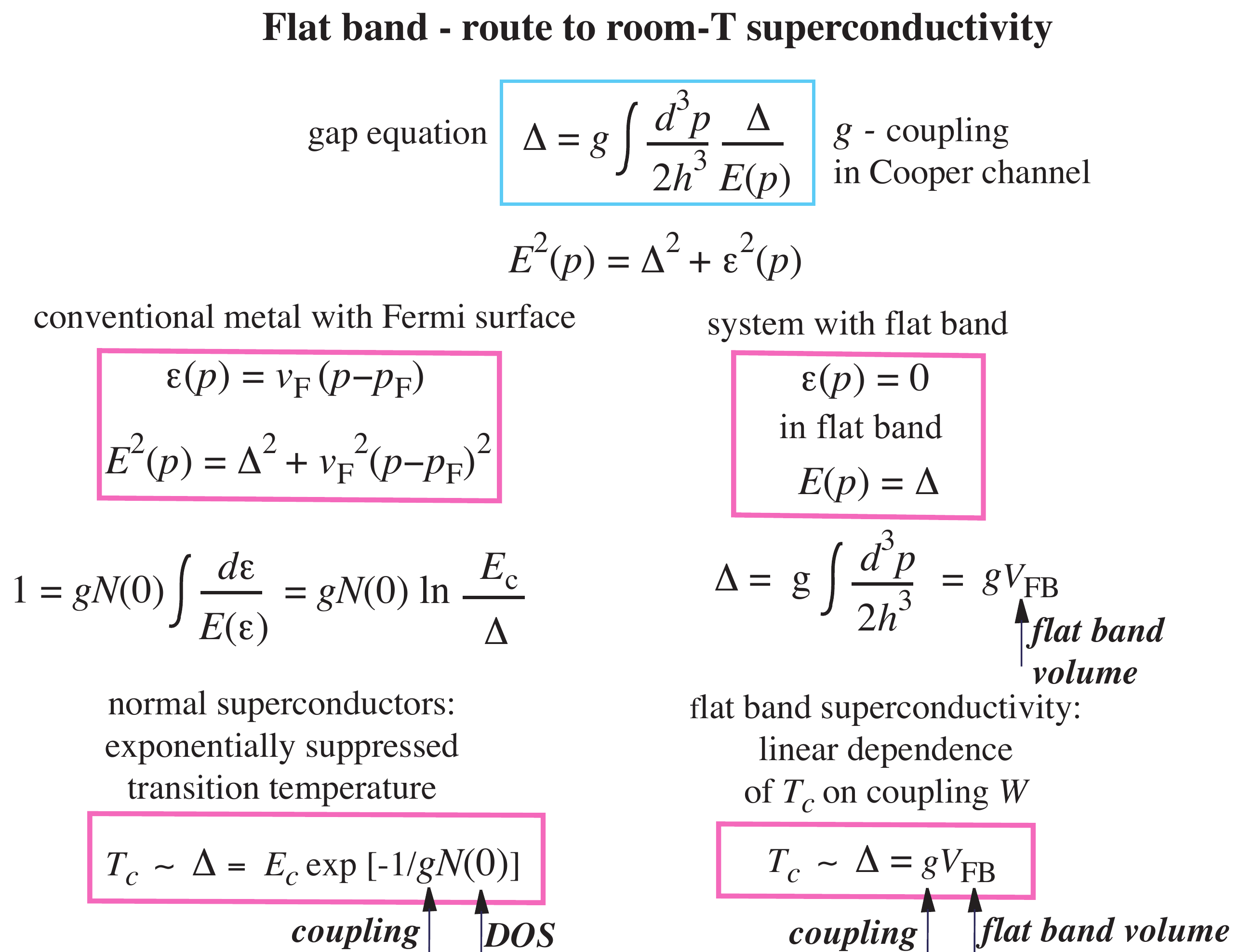}
\caption{Flat band of electronic states with zero energy leads to the linear dependence of transition temperature $T_c$ on the coupling parameter,\cite{Khodel1990} while in conventional metals $T_c$ is exponentially suppressed.
 }
 \label{FlatBandSC}
\end{figure}

The flat band -- or the so-called Khodel-Shaginyan fermion condensate, where all the states have zero energy  -- is caused by electron-electron interaction.\cite{Khodel1990,Volovik1991,Nozieres92} This is the manifestation of the general phenomenon of merging of the energy level due to interaction. Such effect has been observed for Landau levels in 2D quantum wells. \cite{Dolgopolov2014,Dolgopolov2015} Since the flat band has huge density of electronic states, this may considerably enhance the transition temperature to superconducting state, see Fig. \ref{FlatBandSC}.

\begin{figure}
\includegraphics[width=0.5\linewidth]{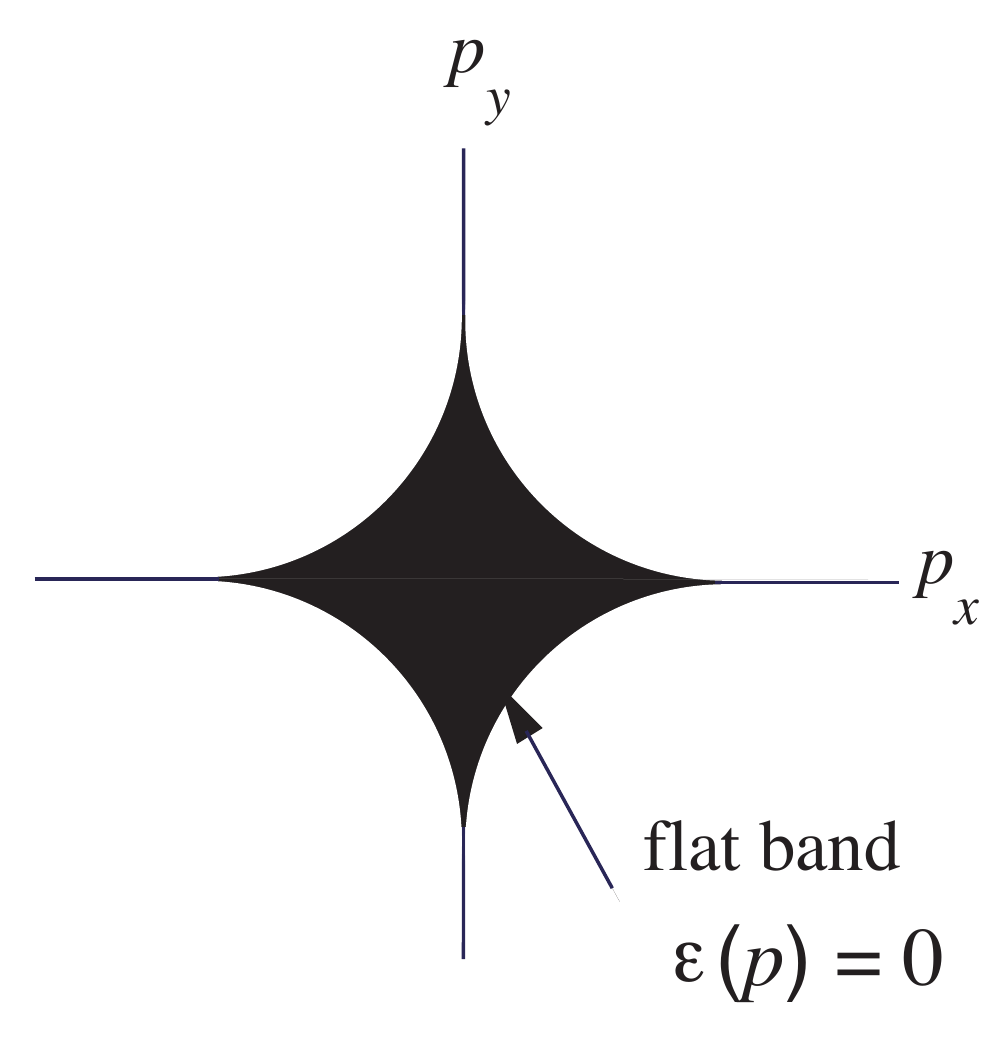}
\caption{Flat band emerging in the vicinity of Lifshitz transition in Fig. \ref{Reconnection} due to electron-electron interaction.\cite{Yudin2014,Volovik1994}  The original non-interacting spectrum is $\epsilon^{(0)}({\bf p})= p_x p_y /m - \mu$, where the Lifshitz transition takes place at $\mu=0$. Due to interaction all the states  in the black region have zero energy (the flat band is shown at the point of Lifshitz transition). 
 }
 \label{FlatSaddle}
\end{figure}

The flat band is more easily formed in the vicinity of the conventional Lifshitz transition,\cite{Yudin2014,Volovik1994}  see Fig. \ref{FlatSaddle}.
Flattening of the single-particle spectrum near the Fermi
momentum has been reported in  2D quantum well.\cite{Dolgopolov2016}
 It is possible that this effect is responsible for the occurrence of superconductivity with high $T$ observed in the pressurized sulfur hydride:\cite{Drozdov2015,Drozdov2016}
there are some theoretical evidences that the high-$T_c$ superconductivity takes place at such pressure, when the system is close to the Lifshitz transition.\cite{Pickett2015,Bianconi2015,Souza2017} 
Enhanced superconductivity at Lifshitz transition has been reported in FeSe monolayer.\cite{FeSe}

\section{Lifshitz transitions governed by Weyl point topology}
\label{Weyl}

\subsection{Topology of Weyl fermions}
\label{WeylTopologySec}

\begin{figure}
\includegraphics[width=1.0\linewidth]{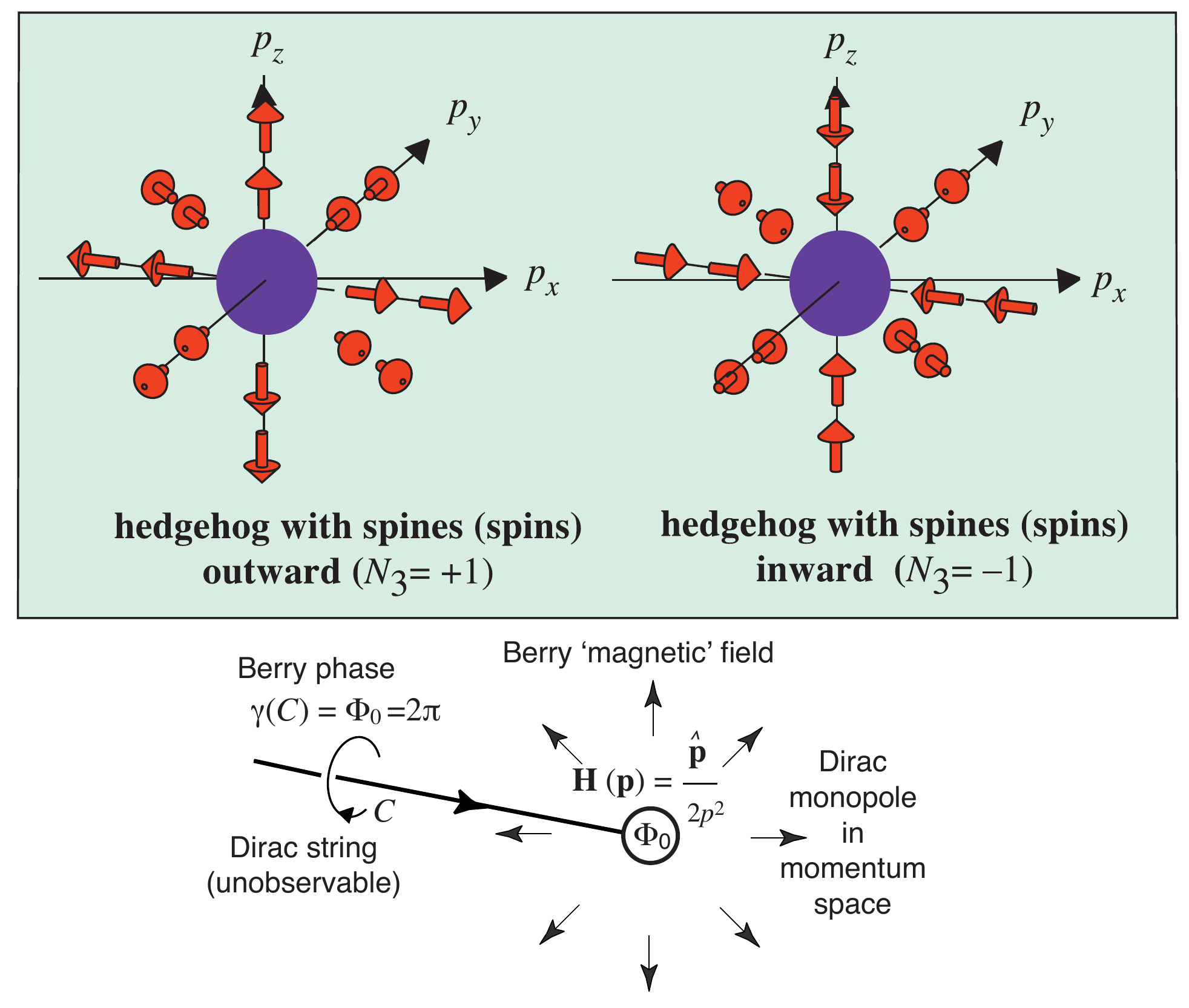}
\caption{Spins of the right-handed Weyl particles (quarks and leptons) are directed along the momentum ${\bf p}$ forming the hedgehog ({\it top left}). The anti-hedgehog -- the hedgehog with spines (spins) inward ({\it top right})  corresponds to the left-handed Weyl particles. The topological invariant $N_3$ describing the topologically distinct hedgehog configurations is expressed either in terms of the Green's function in Eq.(\ref{TopInvariantN3}) or in terms of the unit vector field  in Fig. \ref{objects}   ({\it center}).
 }
 \label{WeylTopology}
\end{figure}

Weyl particles are the elementary particles of our Universe. The Weyl spinor contains 2 complex components, and these massless particles are described by the $2\times 2$ complex Hamiltonian:  $H= c{\mbox{\boldmath$\sigma$}}\cdot{\bf p}$ for right-handed quarks and leptons and $H=- c{\mbox{\boldmath$\sigma$}}\cdot{\bf p}$ for the left-handed particles, where $c$ is the speed of light. Their spins in momentum space form correspondingly the hedgehog and anti-hedgehog in Fig. \ref{WeylTopology}  ({\it top}).
The hedgehog is the topologically stable object, and thus the Weyl point in the center of the hedgehog is topologically protected.
The corresponding topological invariant for the hedgehogs, $N_3$, can be expressed in terms of the Green's function as a surface integral in
the 3+1 momentum-frequency space $p_\mu=({\bf p},\omega)$:
\cite{Volovik2003}
\begin{equation}    
N_3 =\frac {\epsilon_{\mu\nu\rho\sigma}}{24\pi^2}\, 
{\rm tr} \,\oint_{\Sigma_a} dS^{\sigma}  
G\frac{\partial}{\partial p_\mu} G^{-1}\;
G\frac{\partial}{\partial p_\nu} G^{-1}\;
G\frac{\partial}{\partial p_\rho}G^{-1}.
\label{TopInvariantN3}
\end{equation}
Here $\Sigma_a$ is a three-dimensional surface around the
isolated Weyl point in $({\bf p},\omega)$ space.

From the point of view of the general properties of the fermionic spectrum, the Weyl point represents the exceptional point of level crossing  analyzed by von Neumann and Wigner \cite{Neumann1929}. This analysis demonstrates that two branches of spectrum, which have the same symmetry, may touch each other at the conical (or diabolical)  point in the three-dimensional space of parameters, which in our case are $p_x$, $p_y$ and $p_z$. The touching of two branches is described in general by $2\times 2$ Hamiltonian $H= {\mbox{\boldmath$\sigma$}}\cdot{\bf g}({\bf p})$. The topological invariant $N_3$ is expressed in terms of the unit vector 
$\hat{\bf g}({\bf p})= {\bf g}({\bf p})/ | {\bf g}({\bf p})|$ in Fig. \ref{objects}   ({\it center}), which forms the hedgehog configuration in  Fig. \ref{objects}   ({\it middle right}). The touching point also represents the Berry phase monopole in Fig. \ref{WeylTopology}  ({\it bottom}).\cite{Volovik1987} It is not excluded that the Weyl fermions of Standard Model (quark and leptons) are not the elementary particles, but emerge from the level crossing at the more fundamental level.\cite{Froggatt1991,Volovik2003,Horava2005}
In particular, the underlying quantum vacuum can be described by quantum field theory based on real numbers (Majorana fermions), while the imaginary unit, which enters Schr\"odinger equation, emerges in the low energy limit together with the relativistic linear spectrum of Weyl fermions.\cite{VolovikZubkov2014}

The linear ("relativistic") spectrum emerges only for the elementary topological charges, $N_3=+1$ or $N_3=-1$. 
If the Weyl point has higher topological charge, $|N_3|>1$, and if there is no special symmetry, which leads to the degeneracy of the levels, the spectrum  has different dispersion relations along different axes.\cite{VolovikKonyshev1988,Volovik2003}. For example, for $|N_3|=2$ the spectrum is "relativistic" in one direction and quadratic in the other two directions.

\subsection{Lifshitz transition with splitting of Weyl points}
 \label{splitting}

\begin{figure}
\includegraphics[width=1.0\linewidth]{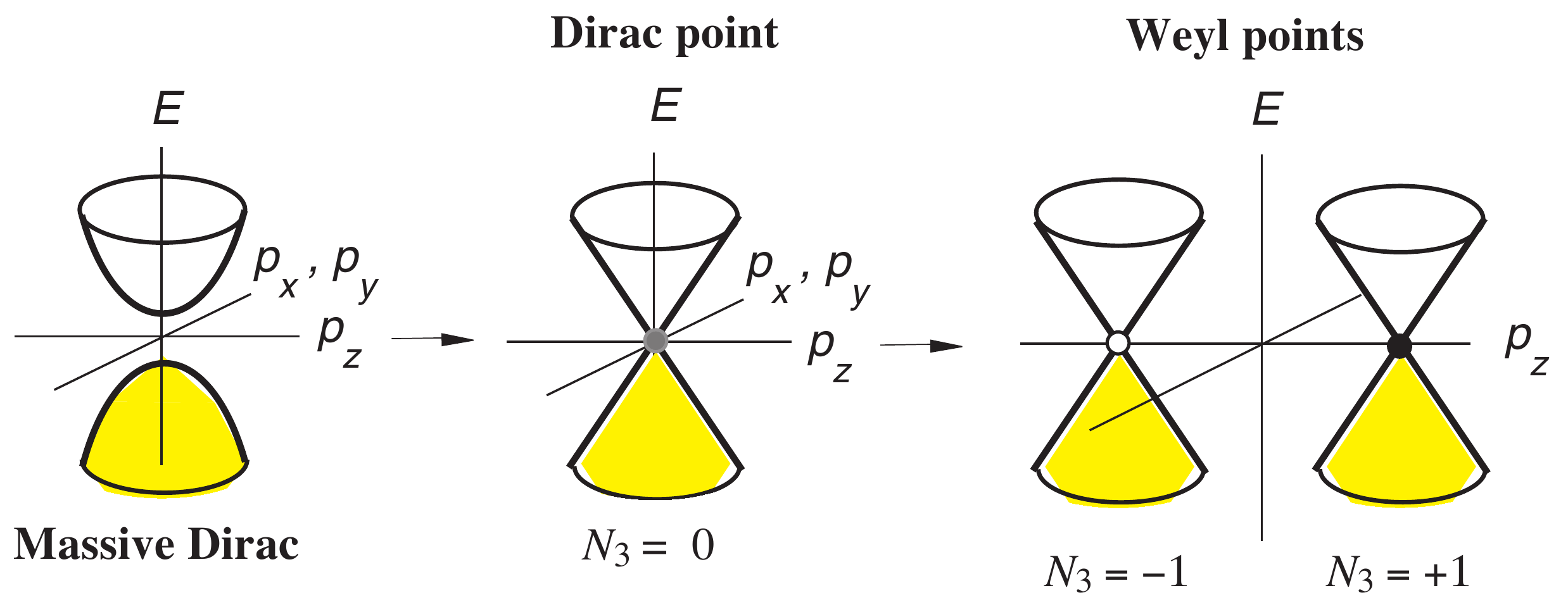}
\caption{Formation of the pair of Weyl points from the vacuum state with massive Dirac fermions.
Topological charges $N_3=+ 1$ and $N_3=- 1$ correspond to right-handed and left-handed particles 
respectively. At the Lifshitz transition the vacuum is gapless with the Dirac point in the fermionic spectrum which has topological charge $N_3=0$.
 }
 \label{DiracToWeyl}
\end{figure}

The typical Lifshitz transition, which involves the Weyl nodes in the fermionic spectrum, describes the formation
of the Weyl points with opposite charges $N_3=\pm 1$ from the fully gapped state.
Fig. \ref{DiracToWeyl} shows the formation of the pair of the Weyl points from the vacuum state with massive Dirac fermions. The intermediate state has the massless Dirac point in the fermionic spectrum with topological charge $N_3=0$. Such gapless Dirac point is marginal, but can be protected by symmetry as this takes place in Standard Model above the electroweak transition. If the symmetry is violated or is spontaneously broken the Dirac spectrum either acquires mass or splits into the pair of Weyl points.\cite{KlinkhamerVolovik2005a}

\begin{figure}
\includegraphics[width=1.0\linewidth]{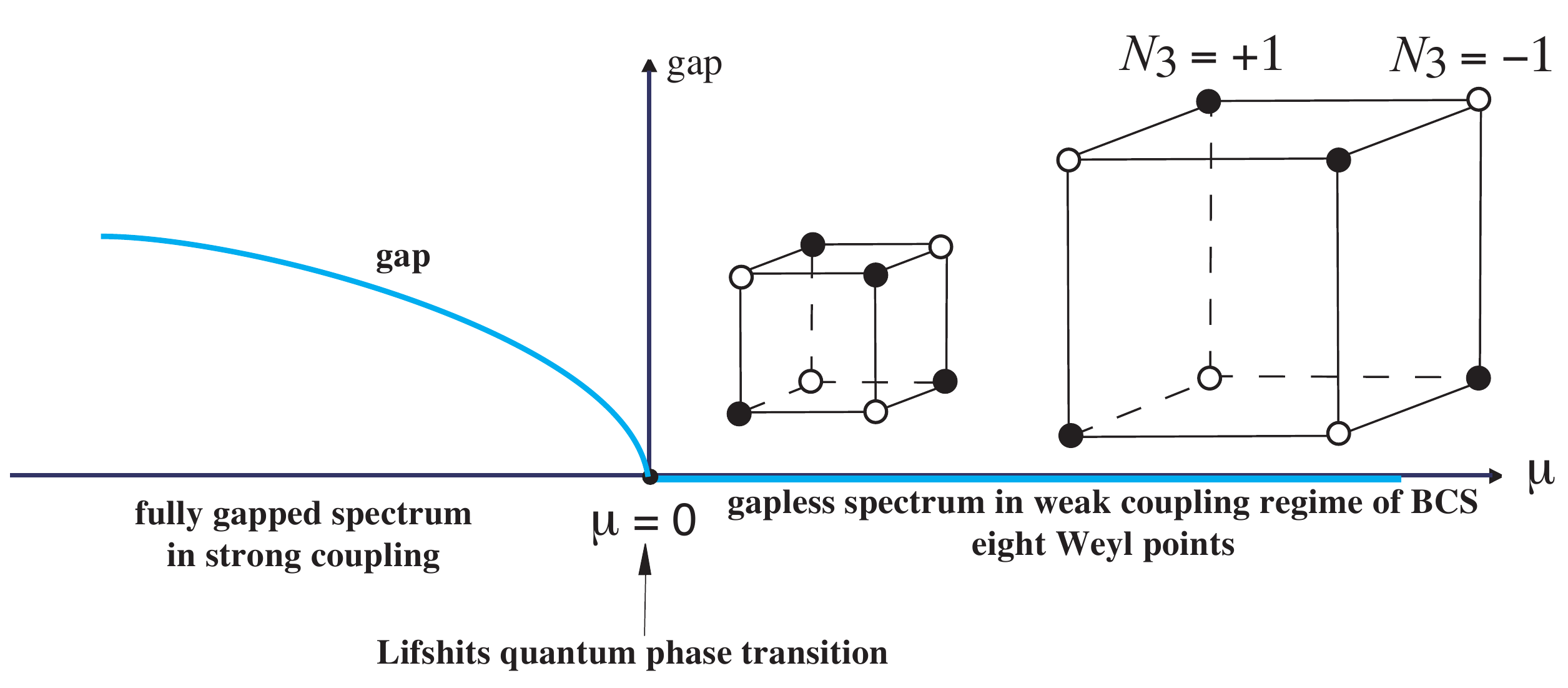}
\caption{The BEC to BCS Lifshitz transition with formation of 4 right-handed and 
4 left-handed Weyl points at vertices of cube. Such arrangement of the Weyl nodes has been discussed for the energy spectrum in superconductors, which belong to the $O(D_2)$ symmetry class.\cite{VolovikGorkov1985,VolovikGorkov2017}
In relativistic theories a simila arrangement gives 8 left and 8 right Weyl fermions on the vertices of the  cube in the 3+1 $(p_x,p_y,p_z,\omega)$ space.\cite{Creutz2008,Creutz2014}
 }
 \label{CubeWeyl}
\end{figure}

Fig. \ref{CubeWeyl} demonstrates the formation of 4 right-handed and 
4 left-handed Weyl points at the Lifshitz transition between the BEC strong coupling regime to the BCS weak coupling regime. Such arrangement of the Weyl nodes takes place in the energy spectrum in the $O(D_2)$ symmetry class of the pair correlated systems.\cite{VolovikGorkov1985,VolovikGorkov2017}
In both cases the total topological charge $N_3({\rm total})=0$, and thus there is an even number of Weyl fermions, which supports the fermion doubling principle.\cite{NielsenNinomiya1981} 
In relativistic theories the analogical arrangement of 8 left and 8 right Weyl fermions on the vertices of the  cube in the 3+1 $(p_x,p_y,p_z,\omega)$ space has been discussed.\cite{Creutz2008,Creutz2014}
It is interesting that each family of Standard Model fermions contains  8 left and 8 right Weyl particles.

\subsection{Lifshitz transition to type-II Weyl cone}
 \label{typeII}

There is the type of Lifshitz transition, which involves both the Fermi surface (invariant $N_1$) and the Weyl point (invariant $N_3$). This is the transition between the isolated Fermi points (type-I Weyl spectrum) to the Weyl point connecting two Fermi surfaces (this is called the type-II Weyl point\cite{Soluyanov2015}). 
In relativistic theories such transitions have been discussed in \cite{VolovikZubkov2014,HuhtalaVolovik2002}.

\begin{figure}
\includegraphics[width=1.0\linewidth]{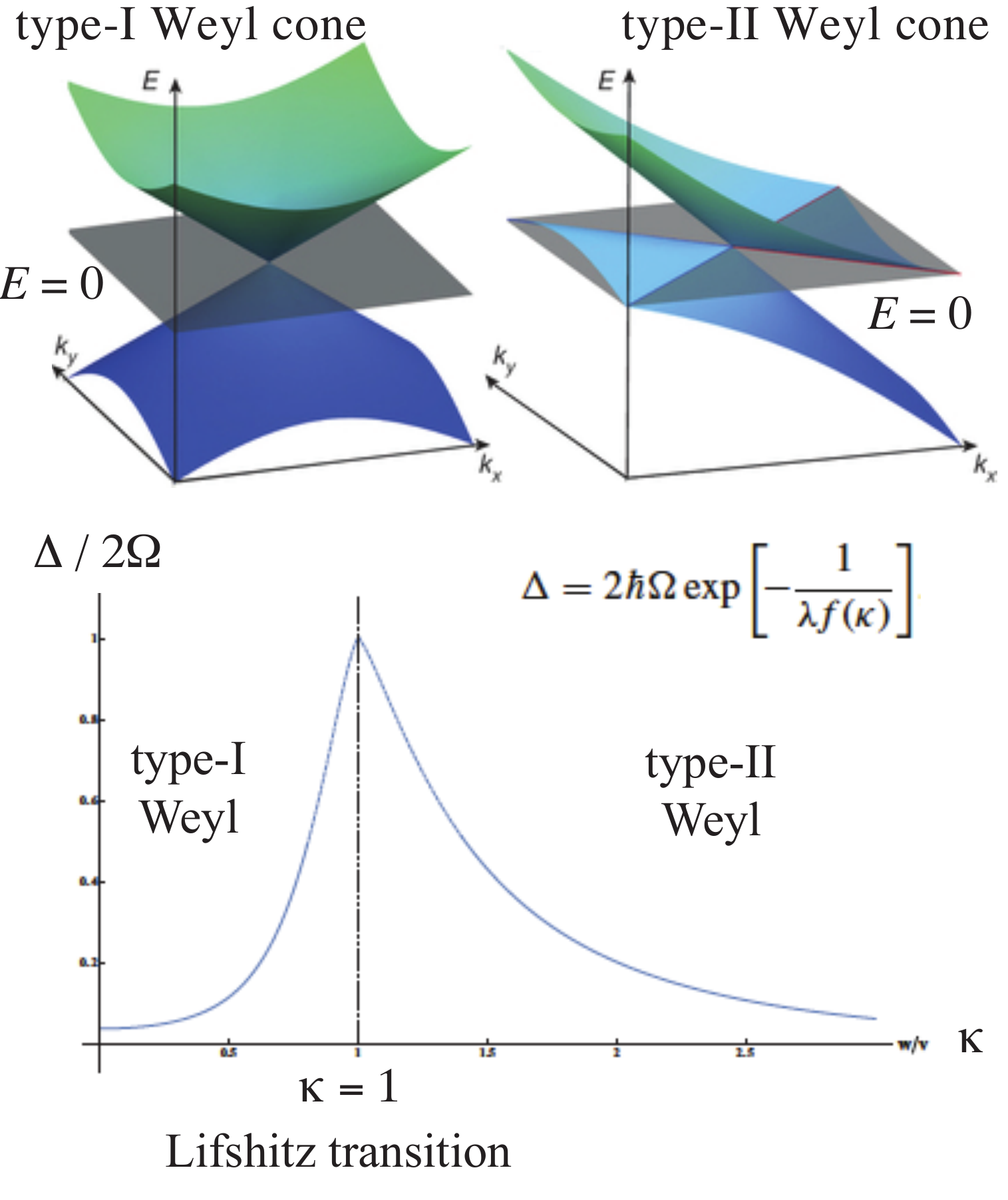}
\caption{({\it top left}): Weyl cone  in the spectrum of type-I Weyl fermions. 
({\it top right}): overtilted Weyl cone in the spectrum of type-II Weyl fermions. ({\it bottom}): Enhancement of superconducting transition temperature $T_c$ at the  Lifshitz transition between the type-I Weyl and the type-II Weyl points.\cite{Shapiro2017}
 }
 \label{WeylType2}
\end{figure}

The simplest realization of the type-II Weyl point comes from the following Hamiltonian with two parameters $c$ and $v$:
\begin{equation}
 H=  
 c{\mbox{\boldmath$\sigma$}} \cdot{\bf p}  - vp_z \,.
 \label{Type2}
 \end{equation}
For $v=0$ this is the Weyl point with the Weyl cone in Fig. \ref{WeylType2} ({\it top left}).
For $0<v<c$ the cone is tilted. At $v>c$ the cone is overtilted, so that the cones cross the zero energy level forming two Fermi pockets connected by the Weyl point --  the type-II Weyl point, see Fig. \ref{WeylType2} ({\it top right}). 
The Lifshitz transition between two types of Weyl point occurs at $v=c$. It is demonstrated that such Lifshitz transition also leads to the enhancement of the transition temperature to 
superconducting state, \cite{Shapiro2017,Zyuzin2017}
see Fig. \ref{WeylType2} ({\it bottom}).

\subsection{Lifshitz transition at the black hole horizon}
 \label{BHhorizon}

\begin{figure}
\includegraphics[width=1.0\linewidth]{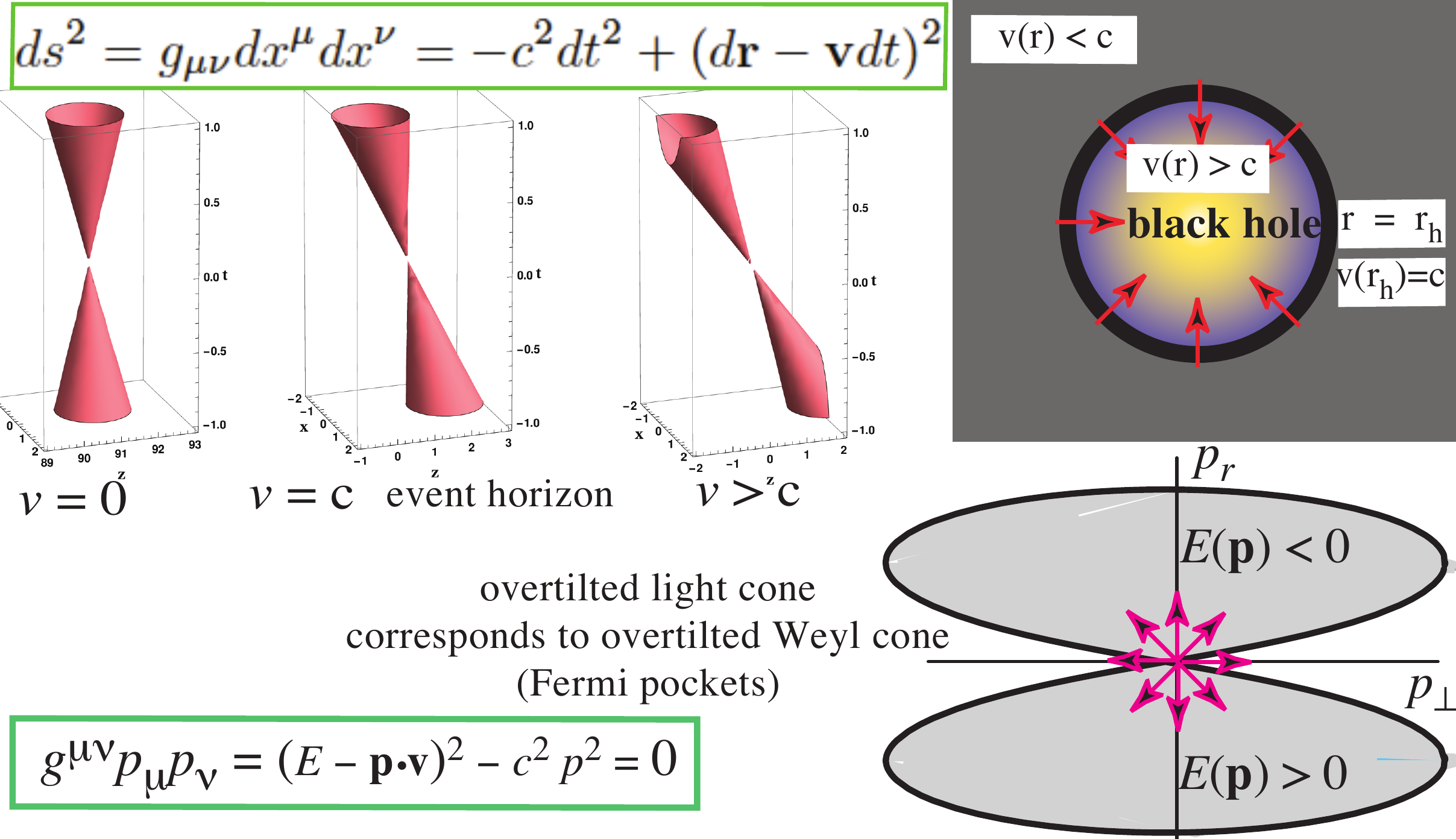}
\caption{Black hole in the Painlev\'e-Gullstrand metric. The metric $g_{\mu\nu}$ describes the light cone. The light cone ({\it top left}) is overtilted behind the horizon, where the frame drag velocity $v>c$,  ({\it top right}).
The metric $g^{\mu\nu}$ describes the Weyl cone. The Weyl cone is overtilted behind the horizon forming two Fermi pockets connected by type-II Weyl point ({\it bottom}). 
The horizon ar $r=r_h$ serves as the surface of the Lifshitz transition between the type-I Weyl point at $r>r_h$ and the type-II Weyl point at $r<r_h$. This behavior of two cones allow us to simulate the black hole horizon and Hawking radiation using Weyl semimetals.\cite{Volovik2016} 
 }
 \label{BH}
\end{figure}

Lifshitz transition discussed in Sec. \ref{WeylType2}
takes place at the black hole horizon. 
 In
general relativity the stationary metric, which is valid both outside and inside the black hole horizon, is provided in particular by the Painlev\'e-Gullstrand spacetime.\cite{Painleve} The line
element of the Painlev\'e-Gullstrand metric is equivalent to the so-called 
acoustic metric:\cite{unruh1,unruh2,Kraus1994} 
\begin{equation}
 ds^2= g_{\mu\nu}dx^\mu dx^\nu=- c^2dt^2+ (d{\bf r}-{\bf v}dt)^2  \,.
\label{Painleve}
\end{equation} 
This metric is expressed in terms of the velocity field ${\bf v}({\bf r})$ 
describing the frame dragging in the gravitational field:
\begin{equation}
{\bf v}({\bf r})= - \hat{\bf r}c\sqrt{\frac{r_h}{r}} ~,~r_h=\frac{2MG}{c^2} \,.
\label{VelocityField}
\end{equation} 
Here $M$ is the mass of the black hole; $r_h$ is the radius of the horizon;
 $G$ is the Newton
gravitational constant.
 Behind the horizon  the drag velocity exceeds the speed of light, $|{\bf v}| > c$, and particles are trapped in the hole, see Fig. \ref{BH} ({\it top right}). The behavior of the light cone  (the cone in spacetime) across the event horizon is shown in Fig. \ref{BH} ({\it top left}). The light cone is overtilted behind the horizon.

The behavior of the Weyl cone (the cone in momentum space) across the horizon is decsribed by Hamiltonian of the Weyl particles in the gravitational field of the black hole, which for the Painlev\'e-Gullstrand metric has the following form:\cite{HuhtalaVolovik2002}
 \begin{equation}
 H=  
\pm c{\mbox{\boldmath$\sigma$}} \cdot{\bf p}  - p_r v(r) \,\,, \,\, v(r)=c\sqrt{\frac{r_h}{r}} \,.
 \label{HamiltonianBH}
 \end{equation}
Here the plus and minus signs correspond to the right handed and left handed Weyl fermions respectively; $p_r$ is the radial component of the linear momentum of the particle. 
Behind the horizon, where  $v>c$ and the light cone is overtilted, the Weyl cone is also overtilted, but in the way  shown in Fig. \ref{WeylType2}. Two Fermi pockets are formed,  which touch each other at type-II  Weyl point in  Fig. \ref{BH} ({\it bottom right}). The event horizon at $r=r_h$ thus serves as the surface of the Lifshitz transition.

 The correspondence between Weyl semimetals and black holes allows us to simulate the black hole horizon
using the inhomogeneous Weyl semimetal, where the transition
between the type-I and type-II Weyl points takes place at some surface.\cite{Volovik2016} This surface would play the role of the event  horizon. The formed black hole will be fully stationary in equilibrium. However, just after creation of this black hole analog, the system is not in the equilibrium state, and the relaxation process at the initial stage of equilibration looks similar to the process of the Hawking radiation.

In the discussed Lifshitz transition between type-I and type-II Weyl points, the element $g_{00}$ of the effective metric changes sign. Another type of Lifshitz transition occurs when the element $g^{00}$ changes sign. In Weyl semimetals this corresponds to the transition to the type-III Weyl fermions,\cite{NissinenVolovik2017} while in general relativity this is the transition to spacetimes with closed timelike curves.

\section{Lifshitz transitions with several topological charges}
\label{DifferentCharges}

\begin{figure}
\includegraphics[width=1.0\linewidth]{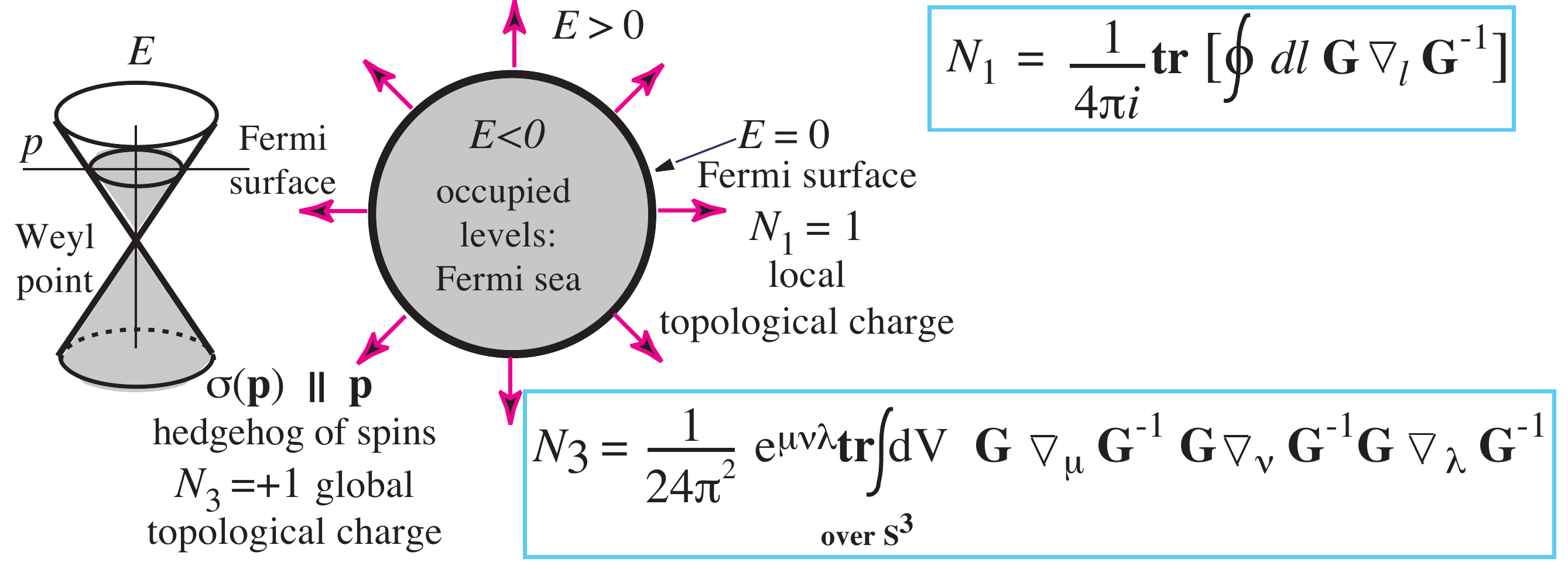}
\caption{Fermi surface with local topological charge $N_1$ and the global topological charge $N_3$.\cite{Volovik2003}
It contains the Berry phase monopole.
 }
 \label{TwoInvariants}
\end{figure}

In sections \ref{typeII} and \ref{BHhorizon} we considered the Lifshitz transition which involved two topological charges:
the charge $N_1$, which characterizes the Fermi surface, and the charge $N_3$ of the Berry phase monopole.
There are the other Lifshitz transitions with the interplay of these two topological invariants. This happens in particular, when the closed Fermi surface is described by two invariants: the local charge $N_1$, which provides the local stability of the Fermi surface, and the global charge $N_3$, which describes the Weyl point inside the Fermi surface  in Fig. \ref{TwoInvariants}. The latter takes place for example when the Weyl point shifts from the zero energy position forming the small Fermi sphere around the Weyl point, see Fig. \ref{TwoInvariants} ({\it left}). Such Fermi sphere contains the $N_3$ charge, which can be obtained from Eq.(\ref{TopInvariantN3}) by integration over the surface, which encloses the Fermi sphere.

\begin{figure}
\includegraphics[width=1.0\linewidth]{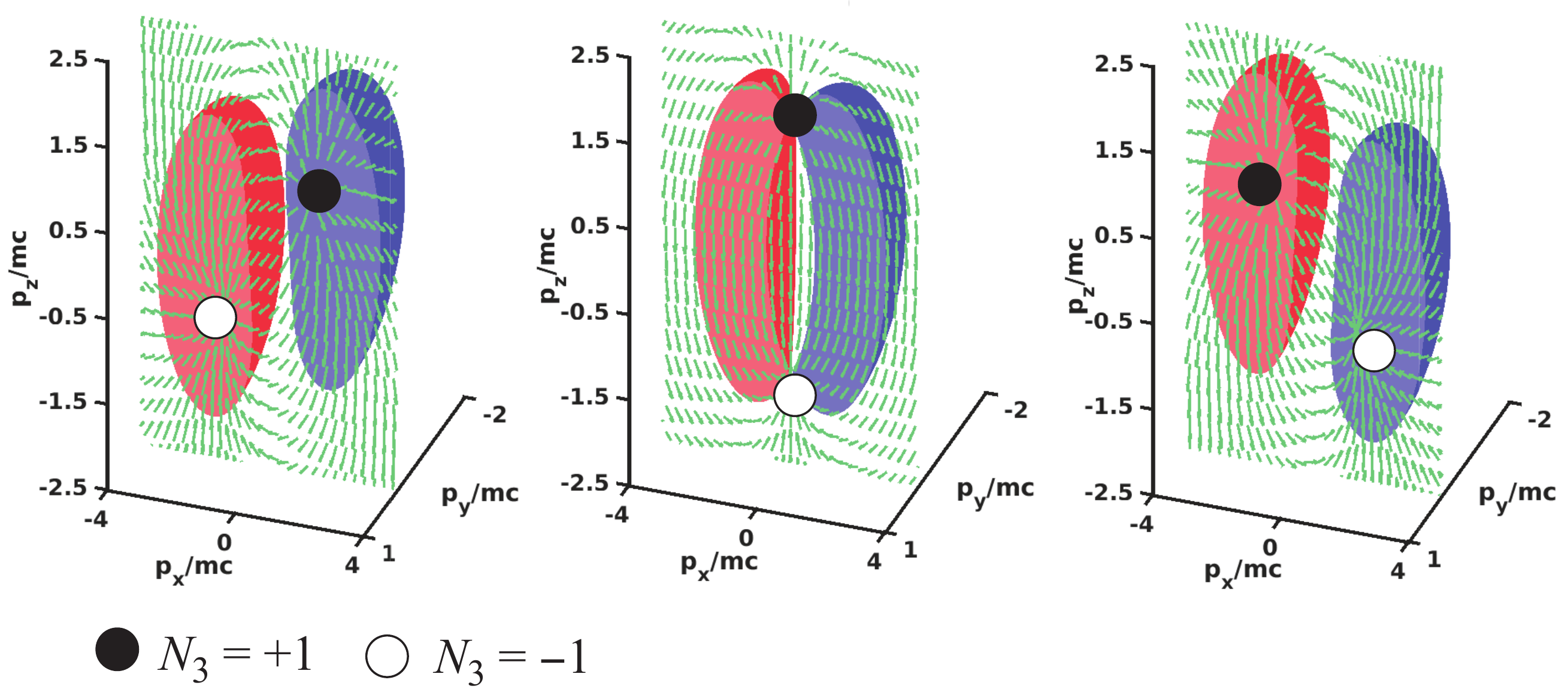}
\caption{Lifshitz transition with exchange of Berry phase monopoles between two Fermi surfaces. \cite{KuangVolovik2017}   The red and  the blue Fermi surfaces are globally non-trivial, with $N_{3}=-1$
and $N_{3}=1$ respectively. In the process of the Lifshitz tramstion, two Berry phase monopoles are pushed out from the  Fermi surfaces. At the Lifshitz transition the Fermi surfaces touch each other at the Weyl points, so that the latter become the type-II Weyl poins. Above the transition the Weyl points are again inside the Fermi surfaces, but now red and  the blue Fermi surfaces have the global charges $N_{3}=+1$
and $N_{3}=-1$ respectively. The topological charges of Fermi surfaces are transferred  between the Fermi surfaces via the type-II Weyl points.
 }
 \label{ExchangeCharge}
\end{figure}

\begin{figure}
\includegraphics[width=1.0\linewidth]{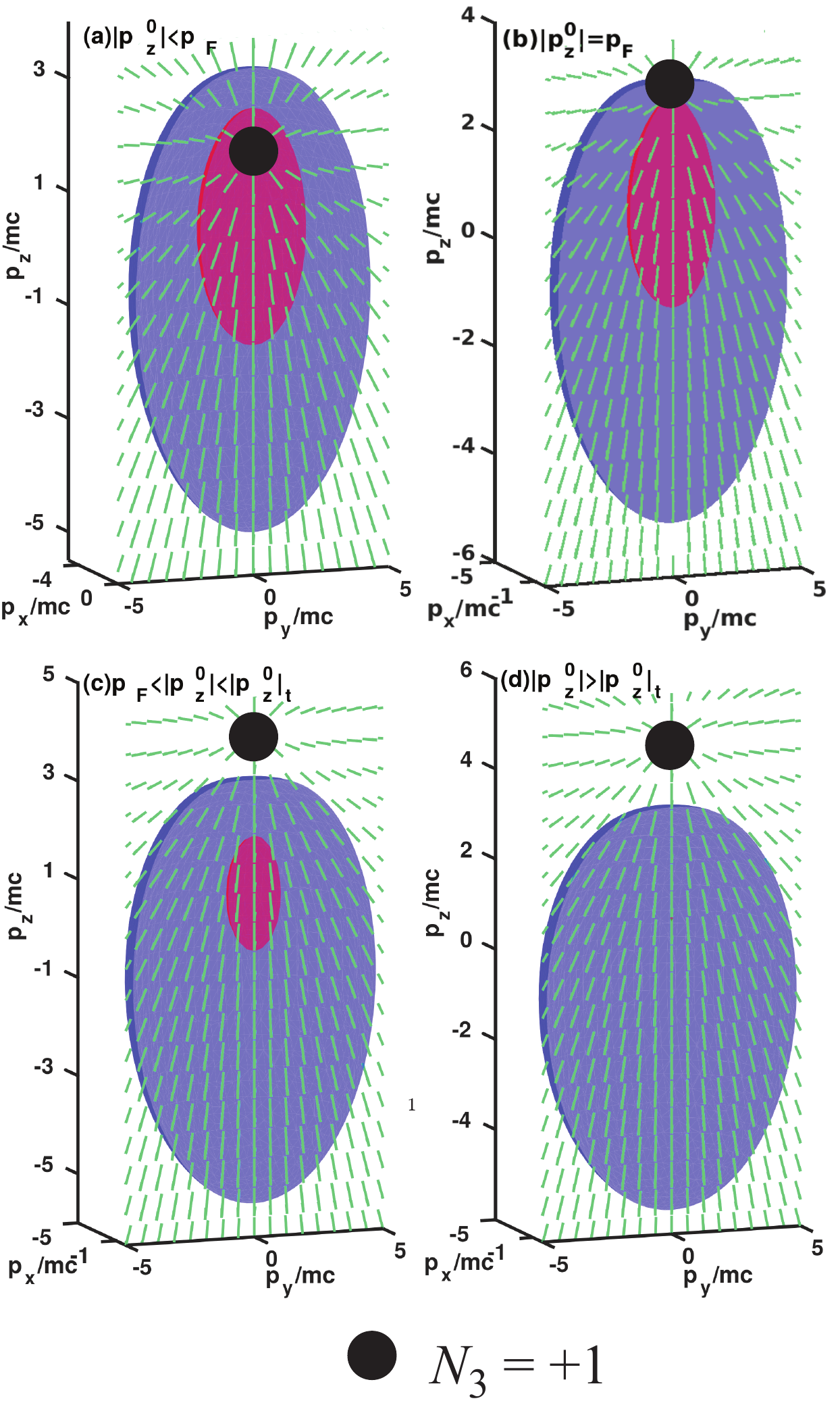}
\caption{  Lifshitz transition, at which the Fermi surfaces loose the Weyl charge 
$N_3$.\cite{KuangVolovik2017} Below the Lifshitz transition both surfaces contain the same Berry phase monopole
with $N_3=+1$. At the  transition the Weyl point connects the inner and outer Fermi surfaces. Above the transition
the monopole comes out from the Fermi surfaces, and both Fermi surfaces become globally trivial, with $N_3=0$. Without global stability the Fermi surface may shrink and disappear in conventional Lifshitz transition as it happens with the red Fermi surface in Fig. \ref{LostCharge} ({\it bottom right}).
 }
 \label{LostCharge}
\end{figure}

At the  Lifshitz transition the Fermi surfaces can exchange their global charges $N_3$ or loose the global charge. \cite{KlinkhamerVolovik2005a,KuangVolovik2017}  
Example of exchange is in Fig. \ref{ExchangeCharge} and the example of the lost global charge is in Fig. \ref{LostCharge}. In both cases the intermediate state at the point of Lifshitz transition contains the type-II Weyl points.

\section{Lifshitz transition governed by conservation of $N_2$ charge}
\label{N2}

The conical Dirac point in the 2D graphene and the nodal lines in the 3D semimetals and nodal superfluids and superconductors are stabilized by the topological charge $N_2$ in Fig. \ref{objects} ({\it bottom}).\cite{Volovik2007,HeikkilaVolovik2015a}
Dirac nodal lines were known to exist in the polar phase of superfluid $^3$He\cite{Dmitriev2015,Autti2016}, in cuprate superconductors,  and in graphite (band crossing lines).\cite{Mikitik2014,Mikitik2006,Mikitik2008} Now they are extensively studied in semimetals.\cite{Takane2017} 

\begin{figure}
\includegraphics[width=1.0\linewidth]{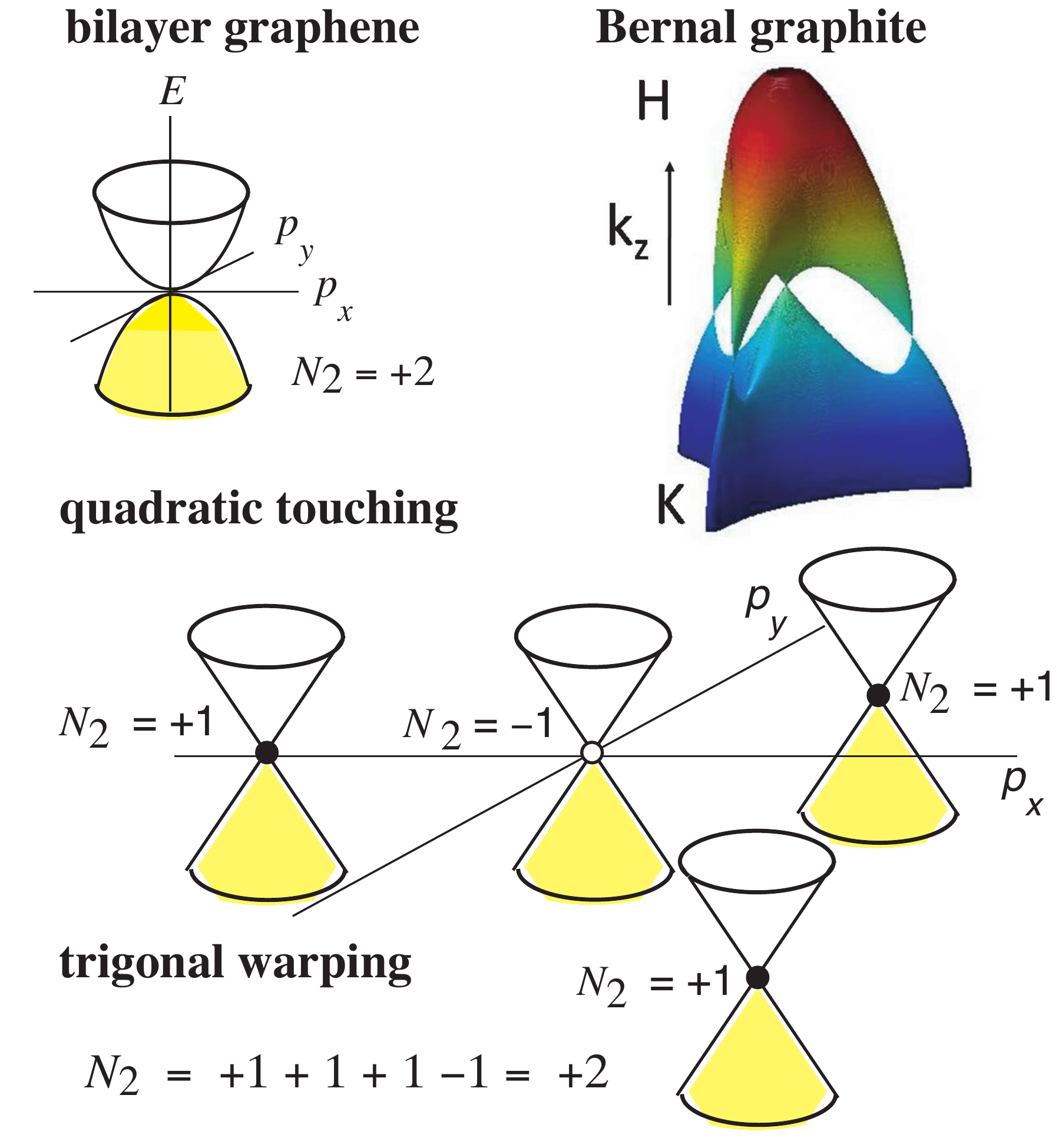}
\caption{
Lifshitz transition governed by conservation of the topological charge  $N_2$ in bilayer graphene.
In bilayer graphene two conical points with the same charge $N_2=1$ on two graphene layers are either merge to form the Dirac point with topological charge $N_2=2$ with quadratic spectrum  in Fig.\ref{Bilayer} ({\it top left}) or split into four Dirac conical points in Fig.\ref{Bilayer} ({\it bottom}). The latter is called the trigonal warping.  The total topological charge $N_2=2$ in both case and thus one configuration may transform to the other one by Lifshitz transition.
 }
 \label{Bilayer}
\end{figure}

The type of Lifshitz transitions governed by the conservation of the topological charge $N_2$ is shown in Fig. \ref{Bilayer} on example of bilayer graphene, when one graphene layer is shifted with respect to the other one. Merging of the two conical points with $N_2=1$ leads to formation of the Dirac node with quadratic dispersion in  Fig. \ref{Bilayer} ({\it top left}), which has the toplogical charge $N_2=2$. This point in turn may split  into four Dirac conical points with $N_2=\pm 1$  in  Fig. \ref{Bilayer} ({\it bottom}). This is the so-called trigonal warping. The total topological charge is conserved, $N_2=1+1+1-1=2$. The trigonal warping can be seen in the Bernal graphite, see  Fig. \ref{Bilayer} ({\it top right}), and the transition occurs as a function of $p_z$,\cite{Mikitik2006,Mikitik2008,Mikitik2014} when $p_z$ crosses the so-called nexus point.\cite{HeikkilaVolovik2015a,HyartHeikkila2016} 

 \section{Lifshitz transitions between gapped states via gapless state}
\label{GappedViaGapless}

\begin{figure}
\includegraphics[width=1.0\linewidth]{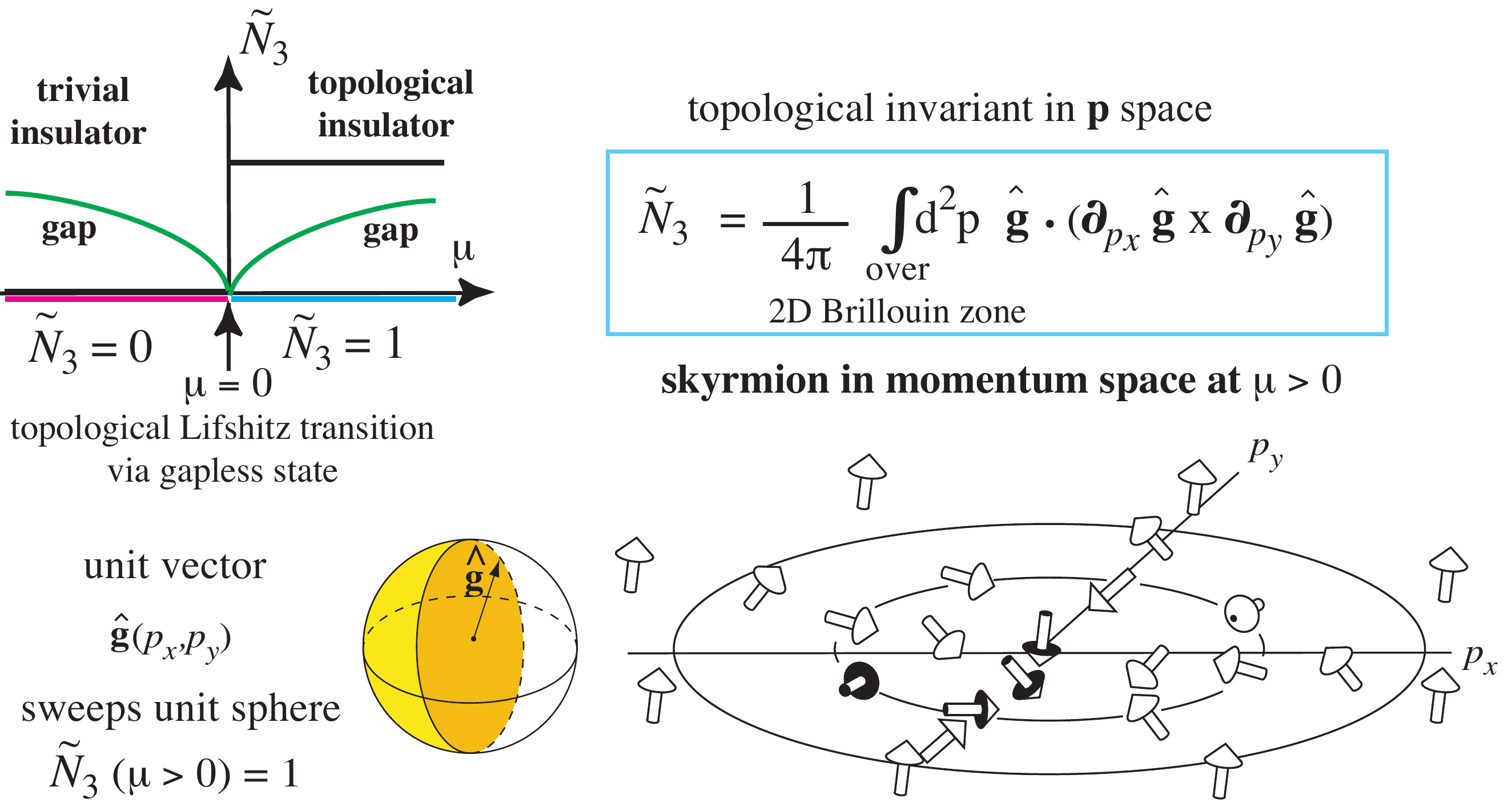}
\caption{Lifshitz transition between the fully gapped topologically different vacua in 2D systems, which experiences the intrinsic quantum Hall 
effect.\cite{So1985,IshikawaMatsuyama1986,IshikawaMatsuyama1987,Haldane1988,Volovik1988} The Hall conductance is expressed in terms of the integer-valued topological invariant  $\tilde N_3$  ({\it top right}). The Lifshitz transition between the topological insulator with  $\tilde N_3=1$ and the trivial insulator with  $\tilde N_3=0$ takes place through the state, where the gap vanishes ({\it top left}).
\\
({\it Bottom}): The  topologically nontrivial state with $\tilde N_3=1$ represents the topologically nontrivial nonsingular object -- skyrmion -- in the 2D momentum space.
 }
 \label{Fully}
\end{figure}

Lifshitz transitions between gapped states include transitions between the topological and non-topological insulators; transitions between the fully gapped superfluids/superconductors;  transitions between the 2D systems, which experience the intrinsic quantum Hall effect; etc. Here we consider such transition on example of the 2D systems, where the Hall conductance is expressed in terms of the integer-valued topological invariant  $\tilde N_3$  in 
Fig. \ref{Fully} ({\it top right}).\cite{So1985,IshikawaMatsuyama1986,IshikawaMatsuyama1987,Haldane1988} 
This topological invariant has the same structure as the invariant $N_3$ in Fig. \ref{objects} ({\it Middle}),
but the integration now is over the whole 2D Brillouin zone. This is an example of the dimensional reduction from the 3D systems with Weyl nodes to the 2D topological insulators.\cite{Volovik2003}

Fig. \ref{Fully} ({\it top left}) demonstrates the Lifshitz transition between the topological insulator with  $\tilde N_3=1$ and the trivial insulator with  $\tilde N_3=0$.
Here the topological charge is not conserved across the Lifshitz transition, but abrubtly changes recalling the first order phase transition. 
Nevertheless the transition occurs smoothly, because at the point of transition the gap in the energy spectrum vanishes. The nullification of the gap at the transition reflects the fact that in the 3D space $(p_x,p_y,\mu)$, where $\mu$ is chemical potential or some other parameter along which the transition occurs, the gap node represents the Weyl point with topological charge 
$N_3=\tilde N_3({\rm right})-\tilde N_3({\rm left})$.\cite{Volovik2003,Kourtis2017}

Example is the 2D $p_x+ip_y$ superfluid/superconductor,\cite{Volovik1988}  where the Lifshitz transition between the superfluid states with $\tilde N_3=1$ and $\tilde N_3=0$ occurs at the same point $\mu=0$ as in the normal Fermi liquid. The detailed consideration shows that the  Lifshitz transition represents the quantum transition of third order:\cite{Ortiz2008} 
the third-order derivative $d^3E/dg^3$ of the
ground state energy $E$ over the interaction strength $g$ is discontinuous. Compare this with  the original  $2\frac{1}{2}$ order transition,\cite{ILifshitz1960} 
and the $3\frac{1}{2}$ order transition discussed recently.\cite{Mikitik2014} 

The nullification of the gap in the fermionic spectrum at the transition between the gapped vacua,  suggests the scenario for the solution of the hierarchy problem: the relativistic quantum vacuum is almost massless because our Universe is very close to the line of the Lifshitz transition.
The reason why nature would prefer the critical line may be that the gapless states on the transition line are able to accommodate more entropy than the gapped states.\cite{Volovik2010}

 \section{Conclusion}

Topological Lifshitz transitions are ubiquitous, since they involve many types of the topological structure of fermionic spectrum:
Fermi surfaces, Dirac lines, Dirac and Weyl points, edge states, Majorana  zero modes, etc. Each of these structures has their own topological invariant, such as $N_1$, $N_2$,  $N_3$,  $\tilde N_3$, etc., which supports the stability of a given class of the topological structure. The topology of the shape of the Fermi surfaces and the Dirac lines, as well as the interconnection of the objects of different dimensionalities in momentum and frequency-momentum spaces, lead to numerous classes of Lifshitz transitions.  

The consequences of Lifshitz transitions are important in different areas of physics. In particular, the singular density of electronic states emerging at the transition is important for the construction of superconductors with enhanced transition temperature; the Lifshitz transition can be in the origin of the small masses of elementary particles in our Universe; the black hole horizon serves as the surface of Lifshitz transition between the vacua with type-I and  type-II Weyl points; etc.

\section*{Acknowledgements}
The work has been supported by the European Research Council
(ERC) under the European Union's Horizon 2020 research and innovation programme (Grant Agreement No. 694248) and  by RSCF (No. 16-42-01100).


\begin{thebibliography}{99}


\bibitem{ILifshitz1960}
I.M. Lifshitz, 
Anomalies of electron characteristics of a metal in the high pressure region, 
Sov. Phys. JETP {\bf 11}, 1130 (1960).

\bibitem{Horava2005}  
P. Ho\v{r}ava,
Stability of Fermi surfaces and $K$-theory,
Phys. Rev. Lett. \textbf{95}, 016405 (2005).

\bibitem{Volovik2003} 
G.E. Volovik, 
{\it The Universe in a Helium Droplet}, 
Clarendon Press,  Oxford (2003).

\bibitem{Volovik2007} 
G.E. Volovik,
Quantum phase transitions from topology in momentum space,
 Springer Lecture Notes in Physics {\bf 718}, 31--73  (2007);
cond-mat/0601372.

\bibitem{Volovik2017} 
G.E. Volovik,
Topological Lifshitz transitions,
Fizika Nizkikh Temperatur {\bf 43}, 57--67 (2017),
arXiv:1606.08318.

\bibitem{KuangVolovik2017} 
Kuang Zhang and G.E. Volovik,
Lifshitz transitions via the type-II Dirac and type-II Weyl points,
Pis'ma ZhETF {\bf 105}, 504--505 (2017), JETP Lett.  {\bf 105},  519--525 (2017).

\bibitem{Volovik1987}
G.E. Volovik, 
Zeros in the fermion spectrum  in superfluid systems as diabolical points,
JETP Lett. {\bf 46}, 98--102 (1987).

\bibitem{Dmitriev2015}
V.V. Dmitriev, A.A. Senin, A.A. Soldatov, and A.N. Yudin, 
Polar phase of superfluid $^3$He in anisotropic aerogel,
Phys. Rev. Lett. {\bf 115}, 165304 (2015).

\bibitem{FrogNielBook} 
C.D. Froggatt   and  H.B. Nielsen,
{\it Origin of Symmetry}, World Scientific, Singapore, 1991.

\bibitem{Volovik2010} 
G.E. Volovik, 
 Topological invariants  for Standard Model: from semi-metal to topological insulator,
 Pis'ma ZhETF {\bf 91}, 61--67 (2010);   JETP Lett. {\bf 91}, 55--61 (2010);
arXiv:0912.0502.

\bibitem{Nielsen2016a}
B.G. Sidharth, A. Das, C.R. Das, L.V. Laperashvili and H.B. Nielsen,
Topological structure of the vacuum, cosmological constant and dark energy,
Int. J. Mod. Phys. A{\bf 31}, 1630051 (2016),
arXiv:1605.01169.

\bibitem{Nielsen2016b}
B.G. Sidharth, A. Das, C.R. Das, L.V. Laperashvili and H.B. Nielsen,
Cosmological constant and the vacuum stability in the Standard Model,
New Advances in Physics {\bf 10}, 1--39 (2016). 

\bibitem{Nielsen2016c}
L.V. Laperashvili, H.B. Nielsen and C.R. Das,
New results at LHC confirming the vacuum stability and Multiple Point Principle,
 Int. J. Mod. Phys. A{\bf 31}, 1650029 (2016).

\bibitem{Nielsen1997}
D.L. Bennett, H.B. Nielsen and C.D. Froggatt, 
Standard model parameters from the multiple point principle and anti-GUT,
arXiv:hep-ph/9710407.

 \bibitem{Volovik2004}
G.E. Volovik, 
Coexistence of different vacua in the effective quantum field theory and multiple point principle, 
JETP Lett. {\bf 79}, 101 (2004),  Pisma ZhETF {\bf 79}, 131 (2004), arXiv:hep-ph/0309144.

 \bibitem{Volovik1989}
G.E. Volovik, 
Zeroes in energy gap in superconductors  with high transition temperature, 
Phys. Lett. A {\bf 142}, 282--284 (1989).

 \bibitem{LiuWilczek2003}
W.V. Liu and F. Wilczek,
Interior gap superfluidity, 
Phys. Rev. Lett. {\bf 90}, 047002 (2003).

 \bibitem{BarzykinGorkov2007}
V. Barzykin and L.P. Gor'kov,
Gapless Fermi surfaces in superconducting CeCoIn$_5$,
Phys. Rev.  B {\bf 76}, 014509 (2007)

 \bibitem{Agterberg2017}
D.F. Agterberg, P.M.R. Brydon, C. Timm,
Bogoliubov Fermi surfaces in superconductors with broken time-reversal symmetry,
Phys. Rev. Lett. {\bf 118}, 127001 (2017).

 \bibitem{Timm2017}
C. Timm, A.P. Schnyder, D.F. Agterberg, and P.M.R. Brydon,
Inflated nodes and surface states in superconducting half-Heusler compounds,
arXiv:1707.02739.

 \bibitem{Volovik1984}
G.E. Volovik,
Superfluid properties of the  A-phase of $^3$He, 
Usp. Fiz. Nauk. {\bf 143}, 73--109;   Soviet Phys.~Usp. {\bf 27}, 363--384.

 \bibitem{Makinen2017}
J.T. M\"akinen, S. Autti, V.B. Eltsov, J. Rysti and G.E. Volovik,
Vortex non-dynamics and exceeding the Landau speed limit in the polar phase of superfluid $^3$He,
 28th International Conference on Low Temperature Physics, LT28, abstract 008, http://www.trippus.se/eventus/userfiles/84948.pdf

\bibitem{VollhardtMakiSchopohl1980} 
D. Vollhardt, K. Maki  and N.  Schopohl,
Anisotropic gap distortion due to superflow and the depairing critical current in  superfluid $^3$He-B,
J. Low Temp. Phys. {\bf 39}, 79--92 (1980).

 \bibitem{Golov2016}
 T. Zhu, M.L. Evans, R.A. Brown, P.M. Walmsley, and A.I. Golov,
Interactions between unidirectional quantized vortex rings,
Phys. Rev. Fluids {\bf 1}, 044502 (2016).

 \bibitem{Voit1995}
J. Voit, 
One-dimensional Fermi liquids,
Reports on Progress in Physics {\bf 58}, 977 (1995),

\bibitem{Dzyaloshinskii2003}  
I. Dzyaloshinskii: 
Some consequences of the Luttinger theorem:   
The Luttinger surfaces in non-Fermi liquids and Mott insulators,
Phys. Rev.  Phys. Rev. B \textbf{68}, 085113 (2003). 


 \bibitem{Farid2009}
B. Farid, A.M. Tsvelik,
Comment on "Breakdown of the Luttinger sum rule within the Mott-Hubbard insulator", by J. Kokalj and P. Prelovsek [Phys. Rev. B {\bf 78}, 153103 (2008), arXiv:0803.4468], 	
arXiv:0909.2886.

 \bibitem{PseudGap}
U.S. Pracht, N. Bachar, L.  Benfatto, G. Deutscher, E. Farber, M. Dressel and M. Scheffler,
Enhanced Cooper pairing versus suppressed phase coherence shaping the superconducting dome
in coupled aluminum nanograins,
Phys. Rev. B {\bf 93}, 100503(R) (2016).


\bibitem{Khodel1990}
V.A. Khodel  and  V.R. Shaginyan,
Superfluidity in system with fermion condensate,
JETP Lett. \textbf{51}, 553 (1990).

\bibitem{Volovik1991}
G.E. Volovik, 
A new class of normal Fermi liquids,
{\it JETP Lett.} \textbf{53}, 222 (1991).

\bibitem{Nozieres92}
P. Nozieres, 
  Properties of Fermi liquids with a finite range interaction,
  J. Phys. (Fr.) {\bf 2}, 443--458 (1992).

 \bibitem{Dolgopolov2014}
A.A. Shashkin, V.T. Dolgopolov, J.W. Clark, V.R. Shaginyan, M.V. Zverev, V.A. Khodel,
Merging of Landau levels in a strongly-interacting two-dimensional electron system in silicon,
Phys. Rev. Lett. {\bf 112}, 186402 (2014).

 \bibitem{Dolgopolov2015}
A.A. Shashkin, V.T. Dolgopolov, J.W. Clark, V.R. Shaginyan, M.V. Zverev, V.A. Khodel,
Interaction-induced merging of Landau levels in an electron system of double quantum wells,
 JETP Letters {\bf 102}, 36 (2015)


\bibitem{Yudin2014}
D. Yudin, D. Hirschmeier, H. Hafermann, O. Eriksson, A.I. Lichtenstein  and M.I. Katsnelson,
Fermi condensation near van Hove singularities within the Hubbard model on the triangular lattice,
Phys. Rev. Lett. {\bf 112}, 070403 (2014).

\bibitem{Volovik1994}
 G.E. Volovik, 
On Fermi condensate: near the saddle point and within the vortex core,
Pis'ma ZhETF {\bf   59}, 798--802  (1994); JETP Lett. {\bf  59}, 830--835 (1994).


 \bibitem{Dolgopolov2016}
M.Yu. Melnikov, A.A. Shashkin, V.T. Dolgopolov, S.-H. Huang, C.W. Liu, S.V. Kravchenko,
Indication of the fermion condensation in a strongly correlated electron system in SiGe/Si/SiGe quantum wells,
 arXiv:1604.08527.


\bibitem{Drozdov2015}
A.P. Drozdov, M.I. Eremets, I.A. Troyan, V. Ksenofontov, S.I. Shylin, 
Conventional superconductivity at 203 K at high pressures, 
Nature {\bf 525}, 73 (2015).

\bibitem{Drozdov2016}
M.I. Eremets and A.P. Drozdov,
High-temperature conventional superconductivity,
Phys. Usp. {\bf 59} 1154--1160 (2016).


\bibitem{Pickett2015}
Yundi Quan and Warren E. Pickett,
Impact of van Hove singularities in the strongly coupled high temperature superconductor H$_3$S,
Phys. Rev. B {\bf 93}, 104526 (2016).

\bibitem{Bianconi2015}
A. Bianconi and T. Jarlborg,
Lifshitz transitions and zero point lattice fluctuations in sulfur hydride showing near
room temperature superconductivity, 
Novel Superconducting Materials {\bf 1}, 37--49 (2015);
arXiv:1507.01093.


\bibitem{Souza2017}
T.X.R. Souza, F. Marsiglio,
The possible role of van Hove singularities in the high T$_c$ of superconducting H$_3$S,
 arXiv:1708.07264.

\bibitem{FeSe} 
X. Shi, Z.-Q. Han, X.-L. Peng, P. Richard, T. Qian, X.-X. Wu, M.-W. Qiu, S.C. Wang, J.P. Hu, Y.-J. Sun, H. Ding,
Enhanced superconductivity accompanying a Lifshitz transition in electron-doped FeSe monolayer,
Nature Communications {\bf 8}, 14988 (2017).


\bibitem{Neumann1929}
J. von Neumann  und E.P. Wigner,
\"Uber das Verhalten von Eigenwerten bei adiabatischen Prozessen,
 Phys. Zeit. {\bf 30}, 467--470 (1929).


\bibitem{Froggatt1991}
C.D. Froggatt   and  H.B. Nielsen,
{\it Origin of Symmetry} 
(World Scientific, Singapore, 1991).

\bibitem{VolovikZubkov2014}  
G.E. Volovik and M.A. Zubkov,
Emergent Weyl spinors in multi-fermion systems,
Nuclear Physics B {\bf 881}, 514--538  (2014).

\bibitem{VolovikKonyshev1988}
G.E. Volovik, V.A. Konyshev,
Properties of the  superfluid systems with multiple zeros in fermion spectrum,
Pisma ZhETF {\bf 47}, 207-- 209 (1988); JETP Lett. {\bf 47}, 250--254 (1988).

\bibitem{KlinkhamerVolovik2005a}
 F.R. Klinkhamer and G.E. Volovik, 
 Emergent CPT violation from the splitting of Fermi points, 
 Int. J.  Mod. Phys. A {\bf 20}, 2795--2812 (2005); 
hep-th/0403037.

\bibitem{VolovikGorkov1985}
G.E. Volovik, L.P.~Gor`kov, 
Superconductivity classes in the heavy fermion systems,
JETP {\bf 61}, 843--854 (1985).

\bibitem{VolovikGorkov2017}
G.E. Volovik,
Dirac and Weyl fermions: from Gor'kov equations to Standard Model (in memory of Lev Petrovich Gorkov),
Pis'ma ZhETF {\bf 105}, 245--246  (2017), JETP Lett.  {\bf 105},  273--277 (2017),
arXiv:1701.01075.

\bibitem{Creutz2008}
M. Creutz,
Four-dimensional graphene and chiral fermions,
JHEP 0804:017 (2008).

\bibitem{Creutz2014}
M. Creutz, 
Emergent spin, 
Annals Phys. {\bf 342},  21--30  (2014).

 \bibitem{NielsenNinomiya1981} 
H.B. Nielsen, M. Ninomiya: 
Absence of neutrinos on a lattice.  I - Proof by homotopy theory, 
Nucl. Phys. B \textbf{185}, 20  (1981); 
Absence of neutrinos on a lattice. II - Intuitive homotopy proof,  
Nucl. Phys. B \textbf{193}, 173 (1981). 

\bibitem{Soluyanov2015}
A.A. Soluyanov, D. Gresch, Zhijun Wang, QuanSheng Wu, M. Troyer, Xi Dai, B.A. Bernevig,  
Type-II Weyl semimetals,
Nature {\bf 527}, 495--498 (2015).

\bibitem{HuhtalaVolovik2002}
P. Huhtala and  G.E. Volovik,  
Fermionic microstates within Painlev\'e-Gullstrand black hole, 
ZhETF {\bf 121}, 995-1003; JETP {\bf 94}, 853-861 (2002); gr-qc/0111055.

\bibitem{Shapiro2017}
Dingping Li, B. Rosenstein, B.Ya. Shapiro, and I. Shapiro,
Effect of the type-I to type-II Weyl semimetal topological transition on superconductivity,
Phys. Rev. B {\bf 95}, 094513 (2017).

\bibitem{Zyuzin2017}
M. Alidoust, K. Halterman, and A. A. Zyuzin,
Superconductivity in type-II Weyl semimetals,
Phys. Rev. B {\bf 95}, 155124 (2017).

\bibitem{Painleve}
 P. Painlev\'e, 
La m\'ecanique classique et la th\'eorie de la relativit\'e, 
C. R. Hebd. Acad. Sci. (Paris) {\bf 173}, 677-680 (1921); 
A. Gullstrand, 
Allgemeine L\"osung des statischen Eink\"orper\-problems in der Einsteinschen Gravitations\-theorie, 
Arkiv. Mat. Astron. Fys. {\bf 16}, 1-15 (1922).

\bibitem{unruh1} 
W.G. Unruh,
Experimental Black-Hole Evaporation, 
 Phys. Rev. Lett. {\bf 46}, 1351 (1981).

\bibitem{unruh2} 
W.G. Unruh,
Sonic analogue of black holes and
the effects of high frequencies on black hole evaporation,
Phys. Rev. D {\bf 51}, 2827--2838 (1995).

\bibitem{Kraus1994} 
P. Kraus and F. Wilczek,
Some applications of a simple stationary line element for the Schwarzschild geometry,
Mod. Phys. Lett. A {\bf 9}, 3713--3719 (1994).

\bibitem{Volovik2016}  
G.E. Volovik,
Black hole and Hawking radiation by type-II Weyl fermions,
Pisma ZhETF {\bf 104}, 660--661 (2016), JETP Lett.  {\bf 104},  645--648 (2016),
arXiv:1610.00521.

\bibitem{NissinenVolovik2017} 
J. Nissinen and G.E. Volovik,
Type-III and IV interacting Weyl points, 
Pisma ZhETF {\bf 105}, 442--443 (2017), JETP Lett.  {\bf 105},  447--452 (2017),
arXiv:1702.04624.

\bibitem{HeikkilaVolovik2015a} 
T.T. Heikkil\"a and G.E. Volovik,
Nexus and Dirac lines in topological materials,
New J. Phys. {\bf 17},  093019 (2015),
arXiv:1505.03277.

\bibitem{Autti2016} 
S. Autti, V.V. Dmitriev, J.T. M\"akinen, A.A. Soldatov, G.E. Volovik,
A.N. Yudin, V.V. Zavjalov, and V.B. Eltsov,
Observation of half-quantum vortices in superfluid $^3$He,
Phys. Rev. Lett. {\bf 117}, 255301 (2016).

\bibitem{Mikitik2014} 
G.P. Mikitik and Yu.V. Sharlai, 
Dirac points of electron energy spectrum, band-contact lines, and electron topological
transitions of $3\frac{1}{2}$
kind in three-dimensional metals,
Phys. Rev. B {\bf 90}, 155122 (2014).

 \bibitem{Mikitik2006} 
G.P. Mikitik and Yu.V. Sharlai, 
Band-contact lines in the electron energy spectrum of graphite,
Phys. Rev. B {\bf 73}, 235112 (2006).

\bibitem{Mikitik2008} 
G.P. Mikitik and Yu.V. Sharlai,
The Berry phase in graphene and graphite multilayers,
Low Temp. Phys. {\bf 34}, 794--780 (2008).

\bibitem{Takane2017} 
D. Takane, K. Nakayama, S. Souma, T. Wada, Y. Okamoto, K. Takenaka, Y. Yamakawa, A. Yamakage, T. Mitsuhashi, K. Horiba, H. Kumigashira, T. Takahashi, and T. Sato,
Observation of Dirac-like energy band and ring-torus
Fermi surface in topological line-node semimetal CaAgAs,
arXiv:1708.06874.

\bibitem{HyartHeikkila2016} 
T. Hyart, T. T. Heikkila,
Momentum-space structure of surface states in a topological semimetal with a nexus point of Dirac lines,
Phys. Rev. B {\bf 93}, 235147 (2016).

\bibitem{So1985} 
H. So,
Induced topological invariants by lattice fermions in odd dimensions,
Prog. Theor. Phys. {\bf 74}, 585--593 (1985).

\bibitem{IshikawaMatsuyama1986} 
K. Ishikawa  and T. Matsuyama,
Magnetic field induced multi component QED in three-dimensions and quantum Hall effect,
Z. Phys. C {\bf 33}, 41--45 (1986). 

\bibitem{IshikawaMatsuyama1987} 
K. Ishikawa and T. Matsuyama,
A microscopic theory of the quantum Hall effect, 
Nucl. Phys. {\bf B~280}, 523--548  (1987).

\bibitem{Haldane1988} 
 F.D.M. Haldane,
 Model for a quantum Hall effect without Landau levels: Condensed-matter realization of the "Parity Anomaly",
Phys. Rev. Lett. {\bf 61}, 2015--2018 (1988).

\bibitem{Volovik1988} 
G.E. Volovik,  
Analog of quantum Hall effect in superfluid $^3$He film,
JETP  {\bf 67}, 1804--1811 (1988).

\bibitem{Ortiz2008} 
S.M.A. Rombouts, J. Dukelsky and G. Ortiz, 
Quantum phase diagram of the integrable $p_x+
ip_y$ fermionic superfluid, 
Phys. Rev. B {\bf 82}, 224510 (2010).

\bibitem{Kourtis2017} 
S. Kourtis, T. Neupert, C. Mudry, M. Sigrist, and Wei Chen,
Weyl-type topological phase transitions in fractional quantum Hall-like systems,
arXiv:1708.04244.


 \end{thebibliography}
\end{document}